\documentclass{elsarticle}

\usepackage{pstricks,pst-all,pst-slpe}
\usepackage{pstricks,pst-node,pst-plot,pst-3d,pst-3dplot,pst-text,pst-eps}
\usepackage{amsmath, amsthm, amssymb, bm}
\usepackage{graphicx}
\usepackage{verbatim} 
\usepackage{etoolbox}
\usepackage[breaklinks=true,colorlinks=true,urlcolor=blue,linkcolor=blue,citecolor=red]{hyperref}

\numberwithin{equation}{section}

\begin{document}

\title{Casimir-Polder energy for axially symmetric systems}

\author[siuc,ntnu]{K. V. Shajesh}
\ead{kvshajesh@gmail.com} 
\ead[url]{http://www.physics.siu.edu/~shajesh}

\address[siuc]{Department of Physics, Southern Illinois University--Carbondale,
Carbondale, Illinois - 62901, USA}
\address[ntnu]{Department of Energy and Process Engineering,
Norwegian University of Science and Technology, N-7491 Trondheim, Norway}

\author[ntnu]{Prachi Parashar} 
\ead{prachi.parashar@ntnu.no}
\ead[url]{https://www.ntnu.edu/employees/prachi.parashar}

\author[ntnu]{Iver Brevik\corref{cor1}}
\cortext[cor1]{Corresponding author}
\ead{iver.h.brevik@ntnu.no} 
\ead[url]{http://folk.ntnu.no/iverhb}

\date{\today}

\begin{abstract}
We develop a formalism suitable for studying Maxwell's equations 
in the presence of a medium that is axially symmetric, in particular
with respect to Casimir-Polder interaction energies. As an
application, we derive the Casimir-Polder interaction energy between
an electric $\delta$-function plate and an anisotropically
polarizable molecule for arbitrary orientations of the principal
axes of polarizabilities of the molecule. We show that in the perfect
conductor limit for the plate the interaction is insensitive to the
orientation of the polarizabilities of the molecule. We obtain the
Casimir-Polder energy between an electric $\delta$-function sphere
and an anisotropically polarizable molecule, again for arbitrary
orientations of the principal axes of polarizabilities of the
molecule. We derive results when the polarizable molecule is either
outside the sphere, or inside the sphere. We present the perfectly
conducting limit for the $\delta$-function sphere, and also the
interaction energy for the special case when the molecule is at the
center of the sphere. Our general proposition is that the
Casimir-Polder energy between a dielectric body with axial symmetry
and an unidirectionally polarizable molecule placed on the axis with
its polarizability parallel to the axis gets non-zero contribution
only from the $m=0$ azimuth mode. This feature in conjunction with
the property that the $m=0$ mode separates into transverse electric
and transverse magnetic modes allows the evaluation of Casimir-Polder
energies for axially symmetric systems in a relatively easy manner.
\end{abstract}

\maketitle
\newpage
\tableofcontents
\newpage
\section{Introduction}

It is well known that in three dimensions the Helmholtz equation
\begin{equation}
\left[ \nabla^2 + \omega^2 \right] \phi({\bf x}) = 0
\label{Helm-def}
\end{equation}
is exactly solvable in only eleven coordinate systems~\cite{Morse:1953mp}.
Specifically, if the surface upon which the boundary conditions
are to be satisfied is not one of these coordinate surfaces,
or if the boundary conditions are not the simple Dirichlet or Neumann type,
the method of separation of variables fails~\cite{Morse:1953mp}.
The static version of the Helmholtz equation is the Laplace equation,
obtained by setting $\omega =0$ in Eq.~(\ref{Helm-def}),
and leads to the plethora of exactly solvable problems in electrostatics.
The vector Helmholtz equation is more stringent and allows exact solutions
only in five coordinate systems \cite{Morse:1953mp}.
A direct consequence of this is that problems in electrodynamics
are often harder to solve even for seemingly simple geometries.

The Maxwell equations that govern classical electrodynamics
in the presence of medium with permittivity and permeability
are tractable whenever the electric and magnetic fields separate 
into transverse electric (TE) and transverse magnetic (TM) modes. 
A class of problems in electrodynamics in the presence of a medium, for 
example, the Casimir effect, often requires the knowledge of the Green dyadic,
which is a lot more unyielding when it comes to separation into 
TE and TM modes.
A medium with planar symmetry and a medium with spherical symmetry
are among the two classic examples in which the Green dyadic,
in the presence of a medium, separate into TE and TM modes.

The electric and magnetic fields, governed by the Maxwell equations
in the presence of a medium, are separable into TE and TM modes for
axially symmetric systems. This statement appears as an exercise
in Chapter VII of Stratton's textbook \cite{Stratton:1941em} in which
it is credited to Max Abraham. However, this does not necessarily imply
that the corresponding Green's dyadic also separates into TE and TM modes in
axially symmetric systems. 
We develop a formalism, closely based on the methods developed 
in Chapter 46 of Ref.~\cite{Schwinger:1998cla}, 
for studying Maxwell's equations in the presence of a medium,
which has the potential of being extended to other coordinate systems.
We study a medium with planar symmetry using rectangular coordinates,
and then again using cylindrical coordinates. We construct the
corresponding rectangular and cylindrical vector eigenfunctions
conducive for investigations of separation of TE and TM modes.
Later, we study a medium with spherical symmetry using spherical
coordinates and construct the spherical vector eigenfunctions.
Our study of these relatively simpler geometries will serve as 
a precursor to a future study of non-trivial geometries with axial symmetry.

Our premise is that the Casimir-Polder interaction energy 
for axially symmetric systems gets non-zero
contributions only from the $m=0$ azimuth mode. Further, the Green dyadic
for the $m=0$ azimuth mode separates into TE and TM modes even though
$m\neq 0$ modes do not necessarily separate into TE and TM modes. 
These observations suggest that the Casimir-Polder energies for 
axially symmetric systems can be calculated fairly easily. 

As an application of the formalism presented here, we calculate the 
Casimir-Polder interaction energy between a polarizable molecule, for 
arbitrary orientations of the principal axes of polarizabilities of the
molecule, and an electric $\delta$-function plate. 
We show that in the perfect conductor limit this interaction is 
insensitive to the orientation of the principal axes of polarizabilities
of the molecule. We repeat the calculation for a polarizable molecule,
again for arbitrary orientations of the principal axes of polarizabilities
of the molecule,
and an electric $\delta$-function sphere, for the case when the molecule is
either inside or outside the sphere.


\section{Maxwell's equations}%

In Heaviside-Lorentz units the monochromatic components of Maxwell's equations,
proportional to $\exp(-i\omega t)$, in the absence of net charges and currents,
and in the presence of electric and magnetic materials with boundaries, are
\begin{subequations}
\begin{align}
{\bm \nabla} \times {\bf E} &= i \omega {\bf B}, \label{MEcrossE} \\
-{\bm \nabla} \times {\bf H} &= i \omega ({\bf D} + {\bf P}).
\label{MEcrossB}
\end{align}%
\label{ME-cross}%
\end{subequations}
These equations imply ${\bm \nabla} \cdot {\bf B}=0$ and
${\bm \nabla} \cdot ({\bf D}+ {\bf P})=0$.
Here ${\bf P}$ is an external source of polarization, in addition
to the polarization of the material in response to the fields 
${\bf E}$ and ${\bf H}$. The external source ${\bf P}$ serves as 
a convenient mathematical tool, and is set to zero in the end.
In the following we neglect non-linear responses and assume that 
the fields ${\bf D}$ and ${\bf B}$ respond linearly to the 
electric and magnetic fields ${\bf E}$ and ${\bf H}$:
\begin{subequations}
\begin{align}
{\bf D}({\bf r},\omega) 
&= {\bm \varepsilon}({\bf r};\omega) \cdot {\bf E}({\bf r},\omega), \\
{\bf B}({\bf r},\omega) 
&= {\bm \mu}({\bf r};\omega) \cdot {\bf H}({\bf r},\omega).
\end{align}%
\label{DB=emuEB}%
\end{subequations}%
Using Eq.\,(\ref{MEcrossE}) in Eq.\,(\ref{MEcrossB}) we construct the following
differential equation for the electric field,
\begin{equation}
\left[ \frac{1}{\omega^2} {\bm\nabla} \times {\bm\mu}^{-1} \cdot {\bm\nabla}
\times \;+{\bf 1} - {\bm\chi} \right] \cdot {\bf E}({\bf r},\omega)
= {\bf P}({\bf r},\omega),
\label{ddE=P}
\end{equation}
where 
\begin{equation}
{\bm\chi}({\bf r};\omega) = {\bm\varepsilon}({\bf r};\omega) - {\bf 1}.
\end{equation} 
The differential equation for the Green's dyadic 
${\bm\Gamma}({\bf r},{\bf r}^\prime)$, that will play a central role 
in the following discussions, is guided by Eq.\,(\ref{ddE=P}),
\begin{equation}
\left[ \frac{1}{\omega^2} {\bm\nabla} \times {\bm\mu}^{-1} \cdot {\bm\nabla}
\times \;+{\bf 1} - {\bm\chi} \right] \cdot 
{\bm\Gamma}({\bf r},{\bf r}^\prime;\omega)
= {\bf 1} \delta^{(3)}({\bf r}-{\bf r}^\prime),
\label{Gd-deq}
\end{equation}
and defines the relation between the electric field and the polarization
source
\begin{equation}
{\bf E}({\bf r};\omega) 
= \int d^3r^\prime {\bm\Gamma}({\bf r},{\bf r}^\prime;\omega)
\cdot {\bf P}({\bf r}^\prime;\omega).
\label{E=GP-def}
\end{equation}
The corresponding dyadic for vacuum, obtained by setting ${\bm\chi}=0$, 
is called the free Green's dyadic and satisfies the equation
\begin{equation}
\left[ \frac{1}{\omega^2} {\bm\nabla} \times {\bm\mu}^{-1} \cdot {\bm\nabla}
\times \;+{\bf 1} \right] \cdot {\bm\Gamma}_0({\bf r},{\bf r}^\prime;\omega)
= {\bf 1} \delta^{(3)}({\bf r}-{\bf r}^\prime).
\label{fGd-deq}
\end{equation}

\section{Casimir-Polder interaction energy: Hypothesis}
\label{sec-cp-prop}

A polarizable molecule, or an atom, or a nanoparticle, positioned at 
${\bf r}_0$, in our model, is suitably described by the dielectric function
\begin{equation}
{\bm\varepsilon}_\text{atom}({\bf r};\omega) - {\bf 1}
= 4\pi {\bm\alpha}(\omega) \delta^{(3)}({\bf r}-{\bf r}_0).
\label{pol-atom-e1}
\end{equation}
In general, for an anisotropically polarizable molecule, we have
the polarizability tensor
\begin{equation}
{\bm\alpha}(\omega) = \alpha_1(\omega) \hat{\bf e}_1 \hat{\bf e}_1
+\alpha_2(\omega) \hat{\bf e}_2 \hat{\bf e}_2
+\alpha_3(\omega) \hat{\bf e}_3 \hat{\bf e}_3,
\label{pmol-eis}
\end{equation}
where $\alpha_i(\omega)$, $i=1,2,3$, are the polarizabilities along
the principal axes of polarizabilities $\hat{\bf e}_i$, which are
the eigenvalues and eigenvectors of the polarizability tensor.
The dispersion in the polarizabilities $\alpha_i(\omega)$ can be modeled
if necessary, though mostly we will not restrict our analysis to any 
specific dispersion model. The Casimir-Polder limit of 
the interaction energy gets contribution from the static limit of 
polarizabilities $\alpha_i(0)$, which is usually independent of any
specific dispersion model.
The Casimir-Polder interaction energy between a polarizable molecule and
a second dielectric body described by the susceptibility tensor 
${\bm\chi}({\bf r};\omega)$,
(restricting our analysis to ${\bm\mu}({\bf r};\omega)={\bf 1}$,)
is given by the expression
\begin{equation}
E = - \int_{-\infty}^\infty d\zeta \int d^3r \, \text{tr}\, 
{\bm\Gamma}({\bf r}_0,{\bf r};\omega) \cdot {\bm\chi}({\bf r};\omega) \cdot 
{\bm\Gamma}_0({\bf r},{\bf r}_0;\omega) \cdot {\bm\alpha}(\omega),
\label{Ecpa}
\end{equation} 
where ${\bm\Gamma}_0$ is the free Green's dyadic satisfying Eq.\,(\ref{fGd-deq})
and ${\bm\Gamma}$ is the Green's dyadic in the presence of
the susceptibility tensor ${\bm\chi}$ satisfying Eq.\,(\ref{Gd-deq}).
Here $\zeta = -i\omega$, is obtained after a Euclidean rotation,
where $\omega$ is a frequency. 

It is sometimes convenient to define 
\begin{equation}
\bar{\bm\Gamma}({\bf r}_1,{\bf r}_2) = \int d^3r \,
{\bm\Gamma}({\bf r}_1,{\bf r};\omega) \cdot {\bm\chi}({\bf r};\omega) \cdot
{\bm\Gamma}_0({\bf r},{\bf r}_2;\omega).
\label{G=GCG0def}
\end{equation}
Using the differential equation for the Green's dyadic in 
Eq.\,(\ref{Gd-deq}) we can deduce the formal relation
$\bar{\bm\Gamma} ={\bm\Gamma} -{\bm\Gamma}_0$.
One can, then, in principle, drop ${\bm\Gamma}_0$ 
to write $\bar{\bm\Gamma} ={\bm\Gamma}$, arguing that 
the dropped term without the information of the second body can not 
contribute to the interaction energy.
This subtraction gives the relatively simpler form for the energy,
$E = - \int d\zeta \,\text{tr}\,
{\bm\Gamma}({\bf r}_0,{\bf r}_0) \cdot {\bm\alpha}(\omega)$. But,
this introduces spurious divergences in the energy, associated to
dropping of ${\bm\Gamma}_0$. 
We shall not replace $\bar{\bm\Gamma} \to {\bm\Gamma}$,
to appreciate the fact that the expression for the interaction energy
in Eq.\,(\ref{Ecpa}), by construction, for disjoint objects,
has no divergences. 

Let us choose the polarizable molecule to be positioned on the 
$z$ axis, ${\bf r}_0=h\,\hat{\bf z}$, and let it be
unidirectionally polarizable only along the $z$ axis, 
corresponding to the choice 
\begin{equation}
{\bm\alpha}(\omega) = \hat{\bf z} \hat{\bf z} \,\alpha(\omega)
\label{mol-pda}
\end{equation}
in Eq.\,(\ref{pmol-eis}).
Let the second dielectric body be described by the susceptibility tensor
\begin{equation}
{\bm\chi}({\bf r};\omega)={\bf 1}_\perp\,{\chi}(\xi;\omega)
\label{chb2-da}
\end{equation}
expressed in terms of coordinates $(\xi,\eta,\phi)$
that satisfy the completeness relations,
\begin{equation}
{\bf 1} = \hat{\bm\phi} \hat{\bm\phi}
+\hat{\bm\eta} \hat{\bm\eta} +\hat{\bm\xi} \hat{\bm\xi},
\qquad {\bf 1}_\perp = \hat{\bm\phi} \hat{\bm\phi}
+\hat{\bm\eta} \hat{\bm\eta},
\end{equation}
where $\phi$ is the azimuth coordinate angle. Let us choose the vector
${\bf s}$ to represent the projection of vector ${\bf r}$ perpendicular
to the azimuth direction $\hat{\bm\phi}$, such that we can write
${\bf r} = ({\bf s},\phi)$.

The dielectric body described by Eq.\,(\ref{chb2-da}) 
and the unidirectionally polarizable molecule of Eq.\,(\ref{mol-pda})
placed on the axis of the dielectric body
constitutes an axially symmetric configuration.
The interaction energy of Eq.\,(\ref{Ecpa}) for this system
involves the volume element $d^3r=d^2s\,\bar sd\phi$,
where $\bar s d\phi$ is the arc length in the direction of ${\bm\phi}$.
The axial symmetry of the configuration requires the dependence
of the Green dyadics in the azimuth angles to have the form $(\phi-\phi_0)$,
see Appendix~\ref{sec-ax-sym-pg}, which implies the 
Fourier transformations for the Green dyadics in Eq.\,(\ref{Ecpa}) to be,
\begin{subequations}
\begin{align}
{\bm\Gamma}({\bf r}_0,{\bf r};\omega) &= \frac{1}{2\pi} \sum_{m=-\infty}^\infty
{\bm\Gamma}_m({\bf s}_0,{\bf s};\omega) \,e^{-im(\phi-\phi_0)}, \\
{\bm\Gamma}_0({\bf r},{\bf r}_0;\omega) &= \frac{1}{2\pi} 
\sum_{m^\prime=-\infty}^\infty 
{\bm\Gamma}_{0\,m}({\bf s},{\bf s}_0;\omega) \,e^{im^\prime(\phi-\phi_0)}.
\end{align}
\end{subequations}
Using these Fourier transformations in the expression for the interaction energy
in Eq.\,(\ref{Ecpa}) we have
\begin{equation}
E = - \int_{-\infty}^\infty d\zeta \alpha(\omega)
\int \frac{{\bar s} d^2s}{2\pi} \chi(\xi;\omega) 
\sum_{m=-\infty}^\infty \hat{\bf z} \cdot 
{\bm\Gamma}_m ({\bf s}_0,{\bf s};\omega) \cdot {\bf 1}_\perp \cdot
{\bm\Gamma}_{0\, m}({\bf s},{\bf s}_0;\omega) \cdot \hat{\bf z}.
\label{Ecpa-as-FT}
\end{equation}
The suggestion seems to be that all the azimuth modes $m$ contribute to
the interaction energy through the sum over $m$ in Eq.\,(\ref{Ecpa-as-FT}).
But, we shall unravel that due to the additional restriction that
position of the molecule ${\bf s}_0$ is on the $z$ axis, 
only the $m=0$ term gives non-zero
contribution to the energy in Eq.\,(\ref{Ecpa-as-FT}).

To elucidate on the proposition that for axially symmetric systems
only the $m=0$ mode contributes to the Casimir-Polder energy,
we shall proceed here to calculate the Casimir-Polder energy for a 
polarizable molecule interacting with a dielectric body 
having planar symmetry. We shall explicate the formalism using
rectangular coordinates, and then again using cylindrical coordinates.
We shall then expound the formalism for the case of a polarizable molecule
interacting with a dielectric body having spherical symmetry.
The transverse electric (TE) and transverse magnetic (TM) modes 
separate for both these geometries, and arguably leads to significant
simplifications that conceals complex interplay of the modes in 
other geometries. But, the simplifications in these geometries
lets us consider more general scenarios than the confines of 
axially symmetric cases, and contrasting these scenarios with the axially
symmetric cases allows us to discover that our hypothesis seems to be true.
We shall present our proposition in Sec.~\ref{sec-thesis}.

\section{Planar systems in rectangular coordinates}%
\label{sec-planeR}

Our primary concern in this article is to elucidate on a formalism
that is useful in the studies of electromagnetic systems with axial
symmetry. A central feature in our formalism involves the construction
of vector eigenfunctions suitable for the problem.
We illustrate this construction for a system with planar symmetry
in rectangular coordinates. To gain insight we shall discuss this
same system again using
cylindrical polar coordinates in Sec.~\ref{sec-plane}.

\begin{figure}[t]
\begin{center}
\includegraphics{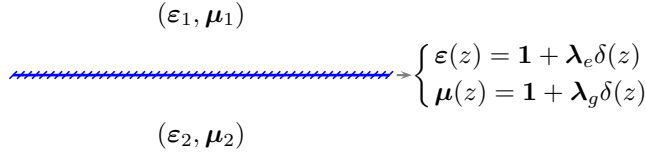}
\caption{A magneto-electric $\delta$-function plate described by properties
$({\bm\lambda}_e,{\bm\lambda}_g)$ sandwiched between
two magneto-electric mediums of infinite extent with properties
$({\bm \varepsilon}_1, {\bm \mu}_1)$ and $({\bm \varepsilon}_2, {\bm \mu}_2)$.}
\label{fig-dplate-b12}
\end{center}
\end{figure}%

We first consider systems with planar symmetry.
As a specific example we consider a $\delta$-function plate sandwiched
between two dielectric mediums of infinite extent,
see Fig.~\ref{fig-dplate-b12}. We choose the direction
of $\hat{\bf z}$ to be in the direction perpendicular to
the plane of $\delta$-function plate.
The gradient operator ${\bm\nabla}$ in rectangular coordinates is
\begin{equation}
{\bm\nabla} = \hat{\bf x} \frac{\partial}{\partial x}
+ \hat{\bf y} \frac{\partial}{\partial y}
+ \hat{\bf z} \frac{\partial}{\partial z}.
\end{equation}
We choose the perpendicular component of the gradient operator to be
\begin{equation}
{\bm\nabla}_\perp = {\bm\nabla} -\hat{\bf z} \frac{\partial}{\partial z}
= \hat{\bf x} \frac{\partial}{\partial x} 
+ \hat{\bf y} \frac{\partial}{\partial y}.
\end{equation}
Further, let us construct the operator
\begin{equation}
\hat{\bf z} \times {\bm\nabla}_\perp = \hat{\bf y} \frac{\partial}{\partial x} 
- \hat{\bf x} \frac{\partial}{\partial y}.
\end{equation}
Given an arbitrary function $f({\bf r})$ the vector field constructions
\begin{equation}
\hat{\bf z} \frac{\partial f}{\partial z}, \quad
{\bm\nabla}_\perp f, \quad (\hat{\bf z} \times {\bm\nabla}_\perp) f,
\end{equation}
are orthogonal to each other at every point in space.
But, this construction is too constrained and not suitable for expressing
electric and magnetic fields, which have three independent components each.
With the goal of expressing the electric and magnetic vector fields, we express
an arbitrary vector field ${\bf E}({\bf r})$ as
\begin{equation}
{\bf E}({\bf r}) = {\bm\nabla}_\perp E^{(u)}({\bf r})
+ (\hat{\bf z} \times {\bm\nabla}_\perp) E^{(v)}({\bf r})
+ \hat{\bf z} E^{(w)}({\bf r}),
\label{ve=thE}
\end{equation}
where $E^{(u)}({\bf r})$, $E^{(v)}({\bf r})$, and $E^{(w)}({\bf r})$,
are arbitrary functions describing the vector field ${\bf E}({\bf r})$.
Observe that
\begin{subequations}
\begin{align}
\big[ \hat{\bf z} E^{(w)}({\bf r}) \big] \cdot
\big[ {\bm\nabla}_\perp E^{(u)}({\bf r}) \big] &= 0, \\
\big[ \hat{\bf z} E^{(w)}({\bf r}) \big] \cdot
\big[ (\hat{\bf z} \times {\bm\nabla}_\perp) E^{(v)}({\bf r}) \big] &= 0, \\
\big[ {\bm\nabla}_\perp E^{(u)}({\bf r}) \big] \cdot 
\big[ (\hat{\bf z} \times {\bm\nabla}_\perp) E^{(v)}({\bf r}) \big] &\neq 0, 
\end{align}%
\label{gr-id-fg}%
\end{subequations}%
which implies that the decomposition of the vector field in Eq.\,(\ref{ve=thE})
is not an orthogonal construction at every point in space. However, we shall
soon be able to construct suitable orthonormal vector eigenfunctions
starting from Eq.\,(\ref{ve=thE}).
To this end, we construct the Laplacian on the surface of a plane
perpendicular to $\hat{\bf z}$,
\begin{equation}
\nabla_\perp^2 = \frac{\partial^2}{\partial x^2}
+ \frac{\partial^2}{\partial y^2}.
\end{equation}
This Laplacian satisfies the eigenvalue equation
\begin{equation}
\nabla_\perp^2 \big[ e^{i {\bf k}_\perp \cdot {\bf r}_\perp} \big]
=-k_\perp^2 \big[ e^{i {\bf k}_\perp \cdot {\bf r}_\perp} \big],
\label{eve=fikx}
\end{equation}
where
\begin{equation}
{\bf k}_\perp \cdot {\bf r}_\perp = k_xx + k_yy, \quad
k_\perp^2 =k_x^2 +k_y^2.
\end{equation}
The eigenfunctions in Eq.\,(\ref{eve=fikx}) are the Fourier eigenfunctions
and satisfy the completeness relation
\begin{equation}
\delta^{(2)}({\bf r}_\perp -{\bf r}_\perp^\prime)
=\int \frac{d^2k_\perp}{(2\pi)^2} e^{i {\bf k}_\perp \cdot {\bf r}_\perp}
e^{-i {\bf k}_\perp \cdot {\bf r}_\perp^\prime}
\end{equation}
and the orthogonality relation
\begin{equation}
\int d^2r_\perp e^{-i {\bf k}_\perp \cdot {\bf r}_\perp}
e^{i {\bf k}_\perp \cdot {\bf r}_\perp^\prime}
=(2\pi)^2\delta^{(2)}({\bf k}_\perp -{\bf k}_\perp^\prime).
\end{equation}

The completeness of the eigenfunctions of the Laplacian on the plane
lets us expand the functions $E^{(i)}({\bf r})$, $i=u,v,w,$
in terms of these eigenfunctions. Thus, we can write
\begin{equation}
E^{(i)}({\bf r})
=\int \frac{d^2k_\perp}{(2\pi)^2} e^{i {\bf k}_\perp \cdot {\bf r}_\perp}
E^{(i)}(z,{\bf k}_\perp), 
\label{fxtok}
\end{equation}
where
\begin{equation}
E^{(i)}(z,{\bf k}_\perp)
=\int d^2x_\perp e^{-i {\bf k}_\perp \cdot {\bf r}_\perp} E^{(i)}({\bf r}). 
\label{fktox}
\end{equation}
Using the Fourier expansions in Eqs.\,(\ref{fxtok}) and (\ref{fktox})
in the general construction proposed in Eq.\,(\ref{ve=thE}) we have
\begin{equation}
{\bf E}({\bf r}) = \int \frac{d^2k_\perp}{(2\pi)^2} 
e^{i{\bf k}_\perp \cdot {\bf r}_\perp}
\left[ i{\bf k}_\perp E^{(u)}(z,{\bf k}_\perp)
+ (\hat{\bf z} \times i{\bf k}_\perp) E^{(v)}(z,{\bf k}_\perp)
+ \hat{\bf z} E^{(w)}(z,{\bf k}_\perp) \right].
\label{Ereck}
\end{equation}

\subsection{Vector eigenfunctions}

We define the Fourier vector eigenfunctions as
\begin{subequations}
\begin{align}
{\bf U}({\bf k}_\perp,{\bf r}_\perp)
&= \frac{1}{ik_\perp}\,{\bm\nabla}_\perp e^{i{\bf k}_\perp \cdot{\bf r}_\perp}
= \hat{\bf k}_\perp e^{i{\bf k}_\perp \cdot {\bf r}_\perp}, \label{uvf-pr} \\
{\bf V}({\bf k}_\perp,{\bf r}_\perp)
&= \frac{1}{ik_\perp} (\hat{\bf z} \times {\bm\nabla}_\perp)\, 
e^{i{\bf k}_\perp \cdot {\bf r}_\perp} = (\hat{\bf z} \times \hat{\bf k}_\perp)
\,e^{i{\bf k}_\perp \cdot {\bf r}_\perp}, \label{vvf-pr} \\
{\bf W}({\bf k}_\perp,{\bf r}_\perp)
&= \hat{\bf z} \,e^{i{\bf k}_\perp \cdot {\bf r}_\perp}.
\end{align}%
\label{Fvef}%
\end{subequations}%
The $ik_\perp$'s in the denominator of the Fourier vector eigenfunctions in
Eqs.\,(\ref{uvf-pr}) and (\ref{vvf-pr}) is interpreted as the eigenvalue of the 
operator $\sqrt{\nabla_\perp^2}$, in the context of Eq.\,(\ref{eve=fikx}). It
is introduced for keep the dimensions of ${\bf U}$, ${\bf V}$, and ${\bf W}$,
the same. Additionally, it serves to normalize the vector eigenfunctions.
We point out that the ${\bf k}_\perp=0$ mode does not exist in the 
construction of ${\bf U}$ and ${\bf V}$,
because the derivative operations return zero for this mode.
This does not cause complications in our considerations
that involve integrations over ${\bf k}_\perp$, because a single point
inside an integral is irrelevant.
In terms of the Fourier vector eigenfunctions of Eqs.\,(\ref{Fvef}) the
decomposition of a vector field in Eq.\,(\ref{Ereck}) takes the form
\begin{align}
{\bf E}({\bf r}) &= \int \frac{d^2k_\perp}{(2\pi)^2} \Big[ 
{\bf U}({\bf k}_\perp,{\bf r}_\perp) E^{(u)}(z,{\bf k}_\perp)
+{\bf V}({\bf k}_\perp,{\bf r}_\perp) E^{(v)}(z,{\bf k}_\perp)
\nonumber \\ & \hspace{20mm}
+{\bf W}({\bf k}_\perp,{\bf r}_\perp) E^{(w)}(z,{\bf k}_\perp) \Big],
\label{Ereckrd}
\end{align}
after the redefinitions,
$ik_\perp E^{(u)}(z,{\bf k}_\perp) \to E^{(u)}(z,{\bf k}_\perp)$
and
$ik_\perp E^{(v)}(z,{\bf k}_\perp) \to E^{(v)}(z,{\bf k}_\perp)$.
Using the notations
\begin{subequations}
\begin{align}
{\bf X}^{(u)}({\bf k}_\perp,{\bf r}_\perp) 
&= {\bf U}({\bf k}_\perp,{\bf r}_\perp), \\
{\bf X}^{(v)}({\bf k}_\perp,{\bf r}_\perp) 
&= {\bf V}({\bf k}_\perp,{\bf r}_\perp), \\
{\bf X}^{(w)}({\bf k}_\perp,{\bf r}_\perp) 
&= {\bf W}({\bf k}_\perp,{\bf r}_\perp),
\end{align}%
\label{FvefX}%
\end{subequations}%
the decomposition of the vector field in the basis of the Fourier vector
eigenfunctions in Eq.\,(\ref{Ereckrd}) can be written in the compact form
\begin{equation}
{\bf E}({\bf r}) = \int \frac{d^2k_\perp}{(2\pi)^2} 
{\bf X}^{(i)}({\bf k}_\perp,{\bf r}_\perp) E^{(i)}(z,{\bf k}_\perp),
\qquad i=u,v,w,
\end{equation}
where we used summation convention in the repeated index $i$, that is,
repeated indices are summed.
The completeness relation of the Fourier vector eigenfunctions 
of Eqs.\,(\ref{Fvef}) is constructed as 
\begin{align}
\delta^{(2)}({\bf r}_\perp -{\bf r}_\perp^\prime)\, {\bf 1}
&= \int \frac{d^2k_\perp}{(2\pi)^2} e^{i{\bf k}_\perp \cdot {\bf r}_\perp}
\Big[ \hat{\bf k}_\perp \hat{\bf k}_\perp
+(\hat{\bf z} \times \hat{\bf k}_\perp) (\hat{\bf z} \times \hat{\bf k}_\perp)
+\hat{\bf z} \hat{\bf z} \Big]
e^{-i{\bf k}_\perp \cdot {\bf r}_\perp} \nonumber \\
&= \int \frac{d^2k_\perp}{(2\pi)^2} {\bf X}^{(i)}({\bf k}_\perp,{\bf r}_\perp)
{\bf X}^{(i)*}({\bf k}_\perp,{\bf r}_\perp^\prime).
\end{align}
The orthogonality of the Fourier vector eigenfunctions is stated as
\begin{equation}
\int d^2r_\perp {\bf X}^{(i)*}({\bf k}_\perp,{\bf r}_\perp)
\cdot {\bf X}^{(j)}({\bf k}_\perp^\prime,{\bf r}_\perp)
= \delta^{ij} (2\pi)^2 \delta^{(2)}({\bf k}_\perp -{\bf k}_\perp^\prime).
\end{equation}

The divergence of a vector field in the basis of
the Fourier vector eigenfunctions of Eqs.\,(\ref{Fvef}) is evaluated as
\begin{equation}
{\bm\nabla}\cdot {\bf E}
= \int \frac{d^2k_\perp}{(2\pi)^2} e^{i{\bf k}_\perp \cdot {\bf r}_\perp}
\left[ ik_\perp E^{(u)}(z,{\bf k}_\perp)
+ \frac{\partial}{\partial z} E^{(w)}(z,{\bf k}_\perp) \right],
\end{equation}
which uses the identities,
\begin{subequations}
\begin{align}
\hat{\bf z} \cdot {\bf U} &=0, & {\bm\nabla}_\perp \cdot {\bf U} &=
ik_\perp e^{i{\bf k}_\perp \cdot {\bf r}_\perp}, &
{\bm\nabla} \cdot {\bf U} &= ik_\perp e^{i{\bf k}_\perp \cdot {\bf r}_\perp}, \\
\hat{\bf z} \cdot {\bf V} &=0, & {\bm\nabla}_\perp \cdot {\bf V} &=0, & 
{\bm\nabla} \cdot {\bf V} &=0, \\ 
\hat{\bf z} \cdot {\bf W} &= e^{i{\bf k}_\perp \cdot {\bf r}_\perp},
& {\bm\nabla}_\perp \cdot {\bf W} &=0, & {\bm\nabla} \cdot {\bf W} 
&= e^{i{\bf k}_\perp \cdot {\bf r}_\perp} \frac{\partial}{\partial z}.
\end{align} 
\end{subequations}
Similarly, the curl of a vector field in the basis of the Fourier
vector eigenfunctions of Eqs.\,(\ref{Fvef}) is evaluated as
\begin{align}
{\bm\nabla}\times {\bf E} &= \int \frac{d^2k_\perp}{(2\pi)^2}
\bigg[ - {\bf U}({\bf k}_\perp,{\bf r}_\perp) \frac{\partial}{\partial z} 
E^{(v)}(z,{\bf k}_\perp) \nonumber \\ & \hspace{15mm}
+ {\bf V}({\bf k}_\perp,{\bf r}_\perp) \left\{ \frac{\partial}{\partial z} 
E^{(u)}(z,{\bf k}_\perp) -ik_\perp E^{(w)}(z,{\bf k}_\perp) \right\}
\nonumber \\ & \hspace{15mm}
+ {\bf W}({\bf k}_\perp,{\bf r}_\perp) 
\,ik_\perp E^{(v)}(z,{\bf k}_\perp) \bigg]
\label{Cu-plR}
\end{align}
using
\begin{subequations}
\begin{align}
\hat{\bf z} \times {\bf U} &= {\bf V}, &
{\bm\nabla}_\perp \times {\bf U} &=0, \\ 
\hat{\bf z} \times {\bf V} &= -{\bf U}, &
{\bm\nabla}_\perp \times {\bf V} &= ik_\perp {\bf W}, \\ 
\hat{\bf z} \times {\bf W} &=0, &
{\bm\nabla}_\perp \times {\bf W} &= -ik_\perp {\bf V},
\end{align}
\end{subequations}
and
\begin{subequations}
\begin{align}
{\bm\nabla} \times {\bf U} &= {\bf V} \frac{\partial}{\partial z}, \\
{\bm\nabla} \times {\bf V} &= ik_\perp {\bf W}
- {\bf U} \frac{\partial}{\partial z}, \\ 
{\bm\nabla} \times {\bf W} &= - ik_\perp {\bf V}.
\end{align}
\end{subequations}

The Maxwell equations in the presence of a medium involves the 
electric and magnetic fields ${\bf E}({\bf r})$ and ${\bf H}({\bf r})$
together with the respective macroscopic
fields ${\bf D}({\bf r})$ and ${\bf B}({\bf r})$. We assume the 
macroscopic fields to be linearly related to the electric and magnetic
fields, given by Eq.\,(\ref{DB=emuEB}). We also require the electric
and magnetic properties of the medium to have planar symmetry.
To this end, we shall restrict our analysis to mediums described by
\begin{subequations}
\begin{align}
{\bm\varepsilon}({\bf r};\omega) &= \varepsilon^\perp(z;\omega) {\bf 1}_\perp
+\varepsilon^{||}(z;\omega) \hat{\bf z} \hat{\bf z}, \\
{\bm\mu}({\bf r};\omega) &= \mu^\perp(z;\omega) {\bf 1}_\perp
+\mu^{||}(z;\omega) \hat{\bf z} \hat{\bf z}.
\end{align}%
\label{emRsym}%
\end{subequations}%
Using the form for the material properties in Eqs.\,(\ref{emRsym})
and the decomposition of a vector field in terms of Fourier vector
eigenfunctions in Eq.\,(\ref{Ereckrd}) together in Eq.\,(\ref{DB=emuEB})
we obtain the following decomposition for the macroscopic fields
in terms of Fourier vector eigenfunctions,
\begin{subequations}
\begin{align}
{\bf D}({\bf r}) &= \int \frac{d^2k_\perp}{(2\pi)^2}
\bigg[ {\bf U}({\bf k}_\perp,{\bf r}_\perp) 
\varepsilon^\perp(z;\omega) E^{(u)}(z,{\bf k}_\perp)
\nonumber \\ & \hspace{5mm}
+{\bf V}({\bf k}_\perp,{\bf r}_\perp) 
\varepsilon^\perp(z;\omega) E^{(v)}(z,{\bf k}_\perp)
+{\bf W}({\bf k}_\perp,{\bf r}_\perp) 
\varepsilon^{||}(z;\omega) E^{(w)}(z,{\bf k}_\perp) \bigg], \\
{\bf B}({\bf r}) &= \int \frac{d^2k_\perp}{(2\pi)^2}
\bigg[ {\bf U}({\bf k}_\perp,{\bf r}_\perp) 
\mu^\perp(z;\omega) H^{(u)}(z,{\bf k}_\perp)
\nonumber \\ & \hspace{5mm}
+{\bf V}({\bf k}_\perp,{\bf r}_\perp) 
\mu^\perp(z;\omega) H^{(v)}(z,{\bf k}_\perp) 
+{\bf W}({\bf k}_\perp,{\bf r}_\perp) 
\mu^{||}(z;\omega) H^{(w)}(z,{\bf k}_\perp) \bigg].
\end{align}%
\label{D=eE-plR}%
\end{subequations}

A dielectric medium with planar symmetry was discussed 
in Ref.~\cite{Parashar:2012it}. A highlight of this discussion was
that we derived the boundary conditions on the electric and magnetic
fields at the interface of two mediums directly from the Maxwell 
equations in first order form, and deduced the boundary conditions
on the related Green's functions unambiguously.  
For convenience we chose $k_y=0$ in the discussion, using the rotational
symmetry of a plane. In our present discussion using the construction
of vector eigenfunctions we can avoid making this choice. Nevertheless,
the first order equations, the boundary conditions, and the related
Green's functions are similar in construction. 
We shall present these discussions using vector eigenfunctions in the 
following subsections.

\subsection{Separation of modes}

We consider a dielectric medium with planar symmetry pictured in 
Fig.~\ref{fig-dplate-b12}. We assume that a $\delta$-function plate
given by
\begin{subequations}
\begin{align}
{\bm\varepsilon}({\bf r};\omega) - {\bf 1}
&= {\bm\lambda}_e(\omega) \delta(z-a), \label{dfun-edef} \\
{\bm\mu}({\bf r};\omega) - {\bf 1} &= {\bm\lambda}_g(\omega) \delta(z-a),
\end{align}
\end{subequations}
resides on the interface of the two dielectric mediums.
Physical interpretation of this model was
presented in Refs.~\cite{Parashar:2012it} and \cite{Milton:2013bm}.
Nevertheless, the results obtained here for the perfect conducting 
cases will be independent of this model.

Using the expression for curl operator in Eq.\,(\ref{Cu-plR}),
in conjunction with Eqs.~(\ref{D=eE-plR}),
the projections of Maxwell's equations in Eqs.~(\ref{ME-cross}) 
along the vector eigenfunctions, ${\bf U}$, ${\bf V}$, and ${\bf W}$, are
\begin{subequations}
\begin{align}
&& \underline{{\bm\nabla} \times {\bf E} = i\omega {\bf B}} & &&&&&
\nonumber \\
{\bf U} &:& -\frac{\partial}{\partial z} E^{(v)}
&= i\omega\mu^\perp H^{(u)}, \label{Max1-U-plR} \\
{\bf V} &:& \frac{\partial}{\partial z} E^{(u)}
-ik_\perp E^{(w)} &= i\omega\mu^\perp H^{(v)}, \label{Max1-V-plR} \\
{\bf W} &:& ik_\perp E^{(v)} &= i\omega\mu^{||} H^{(w)},
\label{Max1-W-plR}
\end{align}%
\label{ME1-plR}%
\end{subequations}
and
\begin{subequations}
\begin{align}
&& \underline{{\bm\nabla} \times {\bf H} = -i\omega ({\bf D} + {\bf P})} & &&&&
\nonumber \\ 
{\bf U} &:& -\frac{\partial}{\partial z} H^{(v)} 
&= -i\omega (\varepsilon^\perp E^{(u)} + P^{(u)}), \label{Max2-U-plR} \\
{\bf V} &:& \frac{\partial}{\partial z} H^{(u)} 
-ik_\perp H^{(w)} &= -i\omega (\varepsilon^\perp E^{(v)} +P^{(v)}), 
\label{Max2-V-plR} \\
{\bf W} &:& ik_\perp H^{(v)} 
&= -i\omega (\varepsilon^{||} E^{(w)} + P^{(w)}). \label{Max2-W-plR}
\end{align}%
\label{ME2-plR}%
\end{subequations}
We have suppressed the variable dependence of the components
of the fields $E^{(i)}(z,{\bf k}_\perp)$, $i=u,v,w,$
in Eqs.~(\ref{ME1-plR}) and (\ref{ME2-plR}).
The boundary conditions satisfied by the electric field components
are derived by integrating Eqs.\,(\ref{ME1-plR}) across the planar 
interface, at $z=a$, to yield
\begin{subequations}
\begin{align}
E^{(u)}(z,{\bf k}_\perp) \Big|^{z=a+}_{z=a-}
&= i\omega \lambda_g^\perp H^{(v)}(a,{\bf k}_\perp), \label{bc1-u-plR} \\
E^{(v)}(z,{\bf k}_\perp) \Big|^{z=a+}_{z=a-}
&= -i\omega \lambda_g^\perp H^{(u)}(a,{\bf k}_\perp), \label{bc1-v-plR} \\
D^{(w)}(z,{\bf k}_\perp) \Big|^{z=a+}_{z=a-}
&= -i{\bf k}_\perp \lambda_e^\perp E^{(u)}(a,{\bf k}_\perp). \label{bc1-w-plR}
\end{align}
\end{subequations}
Similarly, the boundary conditions satisfied by the magnetic field components
are derived by integrating Eqs.\,(\ref{ME2-plR}) across the planar 
interface, at $z=a$, to yield
\begin{subequations}
\begin{align}
H^{(u)}(z,{\bf k}_\perp) \Big|^{z=a+}_{z=a-} 
&= -i\omega \lambda_e^\perp E^{(v)}(a,{\bf k}_\perp), \label{bc2-u-plR} \\
H^{(v)}(z,{\bf k}_\perp) \Big|^{z=a+}_{z=a-} 
&= i\omega \lambda_e^\perp E^{(u)}(a,{\bf k}_\perp), \label{bc2-v-plR} \\
B^{(w)}(z,{\bf k}_\perp) \Big|^{z=a+}_{z=a-} 
&= -i{\bf k}_\perp \lambda_g^\perp H^{(u)}(a,{\bf k}_\perp). \label{bc2-w-plR}
\end{align}%
\end{subequations}%
We also obtain the constraints
\begin{subequations}
\begin{align}
\lambda_e^{||} E^{(w)}(a,{\bf k}_\perp) &= 0, \label{bc1-cons-plR} \\
\lambda_g^{||} H^{(w)}(a,{\bf k}_\perp) &= 0. \label{bc2-cons-plR}
\end{align}%
\label{bc-plR}%
\end{subequations}%
Here, Eqs.\,(\ref{bc1-u-plR}) and (\ref{bc2-u-plR}) 
are obtained by integrating Eqs.\,(\ref{Max1-V-plR}) and (\ref{Max2-V-plR}),
and Eqs.\,(\ref{bc1-v-plR}) and (\ref{bc2-v-plR}) 
are obtained by integrating Eqs.\,(\ref{Max1-U-plR}) and (\ref{Max2-U-plR}).
Eqs.~(\ref{bc1-w-plR}) and (\ref{bc2-w-plR}) are obtained by substituting 
Eqs.~(\ref{Max1-W-plR}) and (\ref{Max2-W-plR}) 
in Eqs.~(\ref{Max1-U-plR}) and (\ref{Max2-U-plR}) respectively
and then integrating across the interface. The constraints in 
Eqs.~(\ref{bc1-cons-plR}) and (\ref{bc2-cons-plR}) are obtained by integrating 
Eqs.~(\ref{Max1-W-plR}) and (\ref{Max2-W-plR}).

We construct electric and magnetic Green's functions
of Refs.~\cite{Schwinger:1977pa, Parashar:2012it}
that satisfy the differential equations,
$g^E$ refers to TE mode and $g^H$ to TM mode
as pointed out in footnote~1 of Ref.~\cite{Parashar:2012it},
\begin{subequations}
\begin{align}
\left[ -\frac{\partial}{\partial z} \frac{1}{\mu^\perp(z)}
\frac{\partial}{\partial z} + \frac{k_\perp^2}{\mu^{||}(z)} 
- \omega^2 \varepsilon^\perp(z)
\right] g^E(z,z^\prime;k_\perp,\omega) &= \delta(z-z^\prime), \\
\left[ -\frac{\partial}{\partial z} \frac{1}{\varepsilon^\perp(z)}
\frac{\partial}{\partial z} + \frac{k_\perp^2}{\varepsilon^{||}(z)} 
- \omega^2 \mu^\perp(z)
\right] g^H(z,z^\prime;k_\perp,\omega) &= \delta(z-z^\prime),
\end{align}%
\label{EM-Gfun}%
\end{subequations}%
and deduce the boundary conditions on these Green's functions,
without having any choice to impose them~\cite{Parashar:2012it}.
In Ref.~\cite{Parashar:2012it} we have shown that all the components
of electric and magnetic fields can be expressed
in terms of the Green's functions of Eqs.\,(\ref{EM-Gfun}).
They can be organized together such that we can write them in terms
of a Green's dyadic of the form in Eq.\,(\ref{E=GP-def}). 
Similarly, the magnetic field can be expressed as 
\begin{align}
{\bf H}({\bf r}) &= \int d^3r^\prime \,
{\bm\Phi}({\bf r},{\bf r}^\prime;\omega) \cdot {\bf P}({\bf r}^\prime), 
\label{H=GP-def}%
\end{align}%
where ${\bm\Gamma}$ and ${\bm\Phi}$ are Green's dyadics.

A vector field is decomposed in the basis of the vector eigenfunctions
as per Eq.\,(\ref{Ereckrd}). A Green's dyadic is a tensor.
Further, the Green's dyadic under consideration has the symmetry of a
plane, and the translational symmetry in the directions
perpendicular to $\hat{\bf z}$ imposes a functional dependence of the 
form ${\bf r}_\perp-{\bf r}_\perp^\prime$. Due to this feature 
their decomposition in the basis of vector eigenfunctions is given
in terms of a single Fourier variable ${\bf k}_\perp$, 
\begin{subequations}
\begin{align}
{\bm\Gamma}({\bf r},{\bf r}^\prime;\omega) &= \int \frac{d^2k_\perp}{(2\pi)^2}
{\bf X}^{(i)}({\bf k}_\perp,{\bf r}_\perp) 
\gamma^{ij}(z,z^\prime;k_\perp,\omega) 
{\bf X}^{(j)*}({\bf k}_\perp,{\bf r}_\perp^\prime), \\
{\bm\Phi}({\bf r},{\bf r}^\prime;\omega) &= \int \frac{d^2k_\perp}{(2\pi)^2}
{\bf X}^{(i)}({\bf k}_\perp,{\bf r}_\perp) \phi^{ij}(z,z^\prime;k_\perp,\omega) 
{\bf X}^{(j)*}({\bf k}_\perp,{\bf r}_\perp^\prime),
\end{align}
\end{subequations}
with the matrix components of the respective dyadics given by
\begin{subequations}
\begin{align}
\gamma^{ij}(z,z^\prime;k_\perp,\omega) &= \left[ \begin{array}{ccc}
\frac{1}{\varepsilon^\perp(z)} \frac{\partial}{\partial z}
\frac{1}{\varepsilon^\perp(z^\prime)} \frac{\partial}{\partial z^\prime}
 g^H & 0 & \frac{1}{\varepsilon^\perp(z)} \frac{\partial}{\partial z}
\frac{ik_\perp}{\varepsilon^{||}(z^\prime)} g^H \\[2mm]
0 & \omega^2 g^E & 0 \\[2mm]
-\frac{ik_\perp}{\varepsilon^{||}(z)} \frac{1}{\varepsilon^\perp(z^\prime)} 
\frac{\partial}{\partial z^\prime} g^H & 0 &
-\frac{ik_\perp}{\varepsilon^{||}(z)}
\frac{ik_\perp}{\varepsilon^{||}(z^\prime)} g^H \end{array} \right],
\label{Gamma=gE}
\end{align}
and
\begin{align}
\phi^{ij}(z,z^\prime;k_\perp,\omega)
&= i\omega \left[ \begin{array}{ccc}
0 & \frac{1}{\mu^\perp(z)} \frac{\partial}{\partial z} g^E
& 0 \\[2mm]
\frac{1}{\varepsilon^\perp(z^\prime)}
\frac{\partial}{\partial z^\prime} g^H & 0 &
\frac{ik_\perp}{\varepsilon^{||}(z^\prime)} g^H \\[2mm]
0 & -\frac{ik_\perp}{\mu^{||}(z)} g^E & 0
\end{array} \right].
\label{Phi=gH}
\end{align}
\end{subequations}
In Eq.\,(\ref{Gamma=gE}) we omitted the term
\begin{equation}
-\delta(z-z^\prime)
\left[ \begin{array}{llr}
\frac{1}{\varepsilon^\perp(z)} \hspace{2mm}& 0 \hspace{2mm} & 0 \\
0 & 0 & 0 \\ 0 & 0 & \frac{1}{\varepsilon^{||}(z)}
\end{array} \right],
\end{equation}
that contains a $\delta$-function and thus never contributes 
to interaction energies between two bodies unless they
are overlapping. We will confine our discussions to disjoint bodies and
drop this term, because it does not contribute.
The Green's dyadics are completely determined in terms of the magnetic
and electric Green's functions that satisfy Eqs.~(\ref{EM-Gfun}),
explicit solutions for which were presented in Ref.~\cite{Parashar:2012it}. 

\subsection{Casimir-Polder interaction energy}

\begin{figure}[t]
\begin{center}
\includegraphics{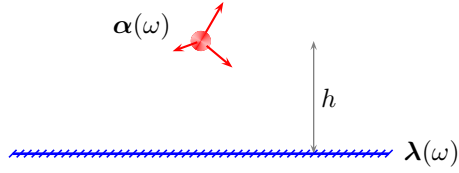}
\caption{An electric $\delta$-function plate described by polarizability
${\bm\lambda}$ interacting with a polarizable molecule of polarizability
${\bm\alpha}(\omega)$, placed height $h$ above the plate.
The three vectors on the molecule illustrate the
directions of the principal axes of polarizabilities,
$\hat{\bf e}_1$, $\hat{\bf e}_2$, and $\hat{\bf e}_3$.}
\label{fig-dplate-atom-CP12}
\end{center}
\end{figure}

We shall now evaluate the interaction energy between a polarizable molecule,
described by Eqs.\,(\ref{pol-atom-e1}) and (\ref{pmol-eis}), and a 
$\delta$-function dielectric plate with a dielectric tensor given by 
Eq.\,(\ref{dfun-edef}) with 
\begin{equation}
{\bm\lambda}_e(\omega) = {\bf 1}_\perp \lambda(\omega),
\quad {\bf 1}_\perp = {\bf 1} -\hat{\bf z}\hat{\bf z}
\label{dp-pol-def}
\end{equation}
and ${\bm\lambda}_g(\omega)=0$.
See Fig.~\ref{fig-dplate-atom-CP12}.
Since the dielectric plate thus constructed 
has translational symmetry in the directions
transverse to $\hat{\bf z}$, the dyadic structure in Eq.\,(\ref{G=GCG0def})
constructed out of ${\bm\Gamma}$, ${\bm\chi}$, and ${\bm\Gamma}_0$,
also has this symmetry. Thus, in the basis of the vector eigenfunctions
of Eqs.\,(\ref{Fvef}) we can write 
\begin{equation}
\bar{\bm\Gamma}({\bf r},{\bf r}^\prime;\omega) 
= \int \frac{d^2k_\perp}{(2\pi)^2} {\bf X}^{(i)}({\bf k}_\perp,{\bf r}_\perp)
\bar\gamma^{ij}(z,z^\prime;k_\perp,\omega)
{\bf X}^{(j)*}({\bf k}_\perp,{\bf r}_\perp^\prime),
\end{equation}
where
\begin{equation}
\bar\gamma^{ij}(z,z^\prime;k_\perp,\omega)
= \gamma^{ii^\prime}(z,a;k_\perp,\omega) \lambda^{i^\prime j^\prime}(\omega)
\gamma_{0}^{j^\prime j}(a,z^\prime;k_\perp,\omega).
\end{equation}
Observe that, due to rotational symmetry, the dependence in ${\bf k}_\perp$
is of the form $k_\perp=|{\bf k}_\perp|$.
We allow the principal axes for the polarizabilities of the molecule
to be arbitrarily oriented while the molecule is at $z=h$ above
the plate.
The expression for Casimir-Polder interaction energy in Eq.\,(\ref{Ecpa})
for this scenario, after using the cyclic property of trace,
can be expressed in the form
\begin{align}
E &= - \int_{-\infty}^\infty d\zeta \int \frac{d^2k_\perp}{(2\pi)^2}
\gamma^{ii^\prime}(h,a;k_\perp,\omega) \lambda^{i^\prime j^\prime}(\omega)
\gamma_{0}^{j^\prime j}(a,h;k_\perp,\omega) \alpha^{ji}(\omega),
\label{cp-mcde-plR}
\end{align}
where, using the shorthand notation 
$\hat{\bf q}_\perp =\hat{\bf z} \times \hat{\bf k}_\perp$,
\begin{align}
\alpha^{ji}(\omega) &= {\bf X}^{(j)*}({\bf k}_\perp,{\bf r}_\perp)
\cdot {\bm\alpha}(\omega) \cdot {\bf X}^{(i)}({\bf k}_\perp,{\bf r}_\perp)
\nonumber \\ &
= \left[ \begin{array}{ccc}
\hat{\bf k}_\perp \cdot {\bm\alpha}(\omega) \cdot \hat{\bf k}_\perp &
\hat{\bf k}_\perp \cdot {\bm\alpha}(\omega) \cdot \hat{\bf q}_\perp &
\hat{\bf k}_\perp \cdot {\bm\alpha}(\omega) \cdot \hat{\bf z} \\
\hat{\bf q}_\perp \cdot {\bm\alpha}(\omega) \cdot \hat{\bf k}_\perp &
\hat{\bf q}_\perp \cdot {\bm\alpha}(\omega) \cdot \hat{\bf q}_\perp &
\hat{\bf q}_\perp \cdot {\bm\alpha}(\omega) \cdot \hat{\bf z} \\
\hat{\bf z} \cdot {\bm\alpha}(\omega) \cdot \hat{\bf k}_\perp &
\hat{\bf z} \cdot {\bm\alpha}(\omega) \cdot \hat{\bf q}_\perp &
\hat{\bf z} \cdot {\bm\alpha}(\omega) \cdot \hat{\bf z} \\
\end{array} \right].
\label{mc-ald12}
\end{align}
Note that the exponential factors inside 
${\bf X}^{(j)*}$'s and ${\bf X}^{(i)}$'s cancel because they are 
evaluated at the same position and are complex conjugates of each other.

Since a $\delta$-function plate does not have longitudinal polarizability,
see Eq.\,(\ref{dp-pol-def}) and discussions in Ref.~\cite{Parashar:2012it},
we have the components describing a $\delta$-function plate given by the matrix
\begin{equation}
\lambda^{ij}(\omega) = \left[ 
\begin{array}{ccc} \lambda(\omega) & 0 & 0 \\
0 & \lambda(\omega) & 0 \\ 0 & 0 & 0 \end{array} \right].
\label{mc-ld12}
\end{equation}
The matrix components of the Green dyadic
in Eq.\,(\ref{Gamma=gE}), for a $\delta$-function plate, are given by 
\begin{equation}
\gamma^{ij}(z,z^\prime;k_\perp,\omega) = \left[ \begin{array}{ccc}
\partial \partial^\prime g^H & 0 & ik_\perp \partial g^H \\[2mm]
0 & \omega^2 g^E & 0 \\[2mm] -ik_\perp \partial^\prime g^H & 0 & k_\perp^2 g^H
\end{array} \right],
\label{mc-gd12}
\end{equation}
and the matrix components of the free Green dyadic
are given by
\begin{equation}
\gamma_0^{ij}(z,z^\prime;k_\perp,\omega)
= \left[ \begin{array}{ccc}
\partial \partial^\prime g_0 & 0 & ik_\perp \partial g_0 \\[2mm]
0 & \omega^2 g_0 & 0 \\[2mm]
-ik_\perp \partial^\prime g_0 & 0 & k_\perp^2 g_0
\end{array} \right],
\label{mc-fgd12}
\end{equation}
where $\partial$ represents derivative with respect to the first variable,
and $\partial^\prime$ represents derivative with respect to
the second variable. The free Green function $g_0(z,z^\prime)$ 
in Eqs.(\ref{mc-gd12}) and (\ref{mc-fgd12}) is given by
\begin{equation}
g_0(z,z^\prime) = \frac{1}{2\kappa} e^{-\kappa |z-z^\prime|}.
\label{g0-pRd}
\end{equation}
The electric and magnetic Green functions for a $\delta$-function plate
are given by
\begin{align}
g^E(z,z^\prime) = \frac{1}{2\kappa} e^{-\kappa |z-z^\prime|}
+r_e^E \frac{1}{2\kappa} e^{-\kappa |z-a|} e^{-\kappa |z^\prime-a|},
\label{gE-pRd}
\end{align}
and
\begin{align}
g^H(z,z^\prime) = \frac{1}{2\kappa} e^{-\kappa |z-z^\prime|}
+ \eta(z-a) \eta(z^\prime-a)
r_e^H \frac{1}{2\kappa} e^{-\kappa |z-a|} e^{-\kappa |z^\prime-a|}, 
\label{gH-pRd}
\end{align}
respectively, which are expressed in terms of the reflection coefficients
\begin{align}
r_e^E = -\frac{\lambda}{\lambda +\frac{2\kappa}{\zeta^2}},
\qquad r_e^H = \frac{\lambda}{\lambda +\frac{2}{\kappa}},
\end{align}
and
\begin{equation}
\kappa^2 = k_\perp^2 +\zeta^2,
\end{equation}
where
\begin{equation}
\eta(z) = \begin{cases} 1, & z>0, \\ -1, & z<0.  \end{cases}
\end{equation}

Using the matrix components for the dyadics in 
Eqs.\, (\ref{mc-ald12}) to
(\ref{mc-fgd12}), in the expression for Casimir-Polder interaction energy 
in Eq.\,(\ref{cp-mcde-plR}) we obtain five terms, 
characterized by 
$\hat{\bf q}_\perp \cdot {\bm\alpha} \cdot \hat{\bf q}_\perp$,
$\hat{\bf k}_\perp \cdot {\bm\alpha} \cdot \hat{\bf k}_\perp$,
$\hat{\bf z} \cdot {\bm\alpha} \cdot \hat{\bf z}$,
$\hat{\bf z} \cdot {\bm\alpha} \cdot \hat{\bf k}_\perp$,
and
$\hat{\bf k}_\perp \cdot {\bm\alpha} \cdot \hat{\bf z}$.
The coefficients of these terms involve derivatives of the Green functions,
given in Eqs.\,(\ref{g0-pRd}), (\ref{gE-pRd}), and (\ref{gH-pRd}),
and we obtain, using Appendix B of Ref.~\cite{Parashar:2012it},
for three of the terms,
\begin{subequations}
\begin{align}
\partial \partial^\prime g^H(h,a) \partial \partial^\prime g_0(a,h)
&= \frac{1}{2\kappa} 
\frac{\kappa^2}{\lambda +\frac{2}{\kappa}} e^{-2\kappa |h-a|}, \\
\omega^4 g^H(h,a) g_0(a,h) &= \frac{1}{2\kappa} 
\frac{\zeta^2}{\lambda +\frac{2\kappa}{\zeta^2}} e^{-2\kappa |h-a|}, \\
k_\perp^2 \partial^\prime g^H(h,a) \partial g_0(a,h) &= \frac{1}{2\kappa} 
\frac{k_\perp^2}{\lambda +\frac{2}{\kappa}} e^{-2\kappa |h-a|}.
\end{align}
\end{subequations}
The remaining two (cross) terms characterized by 
$\hat{\bf z} \cdot {\bm\alpha}(\omega) \cdot \hat{\bf k}_\perp$ and
$\hat{\bf k}_\perp \cdot {\bm\alpha}(\omega) \cdot \hat{\bf z}$
cancel each other, because their coefficients are equal and opposite
in sign, using 
\begin{equation}
\partial \partial^\prime g^H(h,a) \partial g_0(a,h)
=\partial^\prime g^H(h,a) \partial \partial^\prime g_0(a,h)
= -\frac{1}{2} \frac{1}{\lambda +\frac{2}{\kappa}} e^{-2\kappa |h-a|}.
\end{equation}
Thus, we obtain the Casimir-Polder interaction energy between
a polarizable molecule, with polarizability tensor ${\bm\alpha}$, 
distance $h$ above a $\delta$-function plate, choosing $a=0$,
with electric and magnetic reflection coefficients
$r_e^E$ and $r_e^H$, to be 
\begin{align}
E &= - \int \frac{d\zeta d^2k_\perp}{(2\pi)^2}
\frac{e^{-2\kappa h}}{2\kappa}
\Big[\kappa^2 r_e^H \hat{\bf k}_\perp \cdot {\bm\alpha} \cdot \hat{\bf k}_\perp
-\zeta^2 r_e^E \hat{\bf q}_\perp \cdot {\bm\alpha} \cdot \hat{\bf q}_\perp
+k_\perp^2 r_e^H \hat{\bf z} \cdot {\bm\alpha} \cdot \hat{\bf z} \Big],
\label{CP-pl-genE}%
\end{align}
where the integration limits on all three integrals are from $-\infty$
to $\infty$.
A Casimir-Polder interaction energy of a polarizable molecule and a plate
involves orientations of three independent basis vectors:
($\hat{\bf e}_1$, $\hat{\bf e}_2$, $\hat{\bf e}_3$) for the principal
axes of polarizabilities of the molecule;
($\hat{\bf g}_1$, $\hat{\bf g}_2$, $\hat{\bf g}_3$) for the principal
axes of polarizabilities of the plate; and
($\hat{\bf x}$, $\hat{\bf y}$, $\hat{\bf z}$) for the geometrical 
orientation of the plate, which is Fourier transformed in the 
$\hat{\bf x}$ and $\hat{\bf y}$ directions to lead to 
($\hat{\bf k}_\perp$, $\hat{\bf q}_\perp$, $\hat{\bf z}$).
The requirement of planar symmetry for the plane allows the
orientations ($\hat{\bf g}_1$, $\hat{\bf g}_2$, $\hat{\bf g}_3$) 
to be chosen identical to 
($\hat{\bf k}_\perp$, $\hat{\bf q}_\perp$, $\hat{\bf z}$),
which has been used in deriving the expression in Eq.\,(\ref{CP-pl-genE}).
In the presence of planar symmetry for the plane the integrals in
Eq.\,(\ref{CP-pl-genE}) are insensitive to the difference between
$\hat{\bf k}_\perp \cdot {\bm\alpha} \cdot \hat{\bf k}_\perp$ and
$\hat{\bf q}_\perp \cdot {\bm\alpha} \cdot \hat{\bf q}_\perp$. Thus,
inside the integral, we can use
\begin{equation}
\hat{\bf k}_\perp \cdot {\bm\alpha} \cdot \hat{\bf k}_\perp
=\hat{\bf q}_\perp \cdot {\bm\alpha} \cdot \hat{\bf q}_\perp
=\frac{1}{2}\text{tr}\left\{{\bm\alpha}(\omega) \cdot {\bf 1}_\perp\right\},
\qquad {\bf 1}_\perp = {\bf 1} - \hat{\bf z} \hat{\bf z}.
\end{equation}
Using this feature of planar symmetry in Eq.\,(\ref{CP-pl-genE}) we have
\begin{align}
E &= -\frac{1}{4\pi} \int_{-\infty}^\infty d\zeta \int_0^\infty k_\perp dk_\perp
\frac{e^{-2\kappa h}}{2\kappa}
\nonumber \\ & \hspace{5mm} \times
\Big[ (\kappa^2 r_e^H -\zeta^2 r_e^E)\, \text{tr}\,{\bm\alpha}(\omega)
+(r_e^H \kappa^2 -2r_e^H \zeta^2 +r_e^E \zeta^2)\,
\hat{\bf z} \cdot {\bm\alpha}(\omega) \cdot \hat{\bf z} \Big].
\label{CP-pl-genEan}%
\end{align}
The energy of Eq.\,(\ref{CP-pl-genEan}) depends on the orientations of
the principal axes of polarizabilities of the molecule completely
through the term containing 
$\hat{\bf z} \cdot {\bm\alpha}(\omega) \cdot \hat{\bf z}$,
because $\text{tr}{\bm\alpha}(\omega)$ is independent of these orientations. 
We shall not explore the expression in Eq.\,(\ref{CP-pl-genEan})
in its full generality.
A counterpart of the expression in Eq.\,(\ref{CP-pl-genEan}),
for the case of a dielectric slab of finite thickness, was reported and
analyzed in Ref.~\cite{Thiyam:2015fts}.

Let us now consider the $\delta$-function plate to be a perfect conductor.
The energy for this case is obtained by taking the limit 
$\lambda\to\infty$, entailing $r_e^H\to 1$ and $r_e^E\to -1$, in
the expression for energy in Eq.\,(\ref{CP-pl-genEan}). Further,
in the Casimir-Polder limit only the low frequency fluctuations are
dominant, which is obtained in the static limit 
${\bm\alpha}(\omega) \to {\bm\alpha}(0)$. In these ideal limits
all the integrals in Eq.\,(\ref{CP-pl-genEan}) can be completed to yield
\begin{equation}
E=-\frac{\text{tr}\,{\bm\alpha}(0)}{8\pi h^4}.
\label{CP-rCl}
\end{equation}
It is remarkable, and not a priori intuitive,
that in the static and perfect conductor limit the interaction
energy should be insensitive to the orientation of the polarizabilities.
We emphasize that translational symmetry and static polarizability 
alone is not sufficient to render the energy independent of the orientation. 
It is in fact the fine cancellations
in the contributions from the electric and magnetic modes
in the perfect conductor limit that makes the energy orientation 
independent. The analogous result of Eq.\,(\ref{CP-rCl}), for 
isotropic polarizabilities, is the well known result by 
Casimir and Polder, first derived in 1947~\cite{Casmir:1947hx}.

At the loss of considerable generality, but to gain insight from 
the substantial simplicity that follows, we narrow down our attention 
to the case of a unidirectionally polarizable atom polarizable 
in the direction of the axis, described by
\begin{equation}
{\bm\alpha}(\omega) = \alpha(\omega) \, \hat{\bf z} \hat{\bf z}.
\end{equation}
For this case Eq.\,(\ref{CP-pl-genEan}) takes the form
\begin{equation}
E = -\frac{1}{2\pi} \int_{-\infty}^\infty d\zeta \int_0^\infty k_\perp dk_\perp
\frac{e^{-2\kappa h}}{2\kappa} k_\perp^2 r_e^H \alpha(\omega).
\end{equation}
This is the scenario we are focussing on in this article, because it is
an axially symmetric system. If we further assume the static
and perfect conductor limit, we can complete the integrals to obtain
\begin{equation}
E=-\frac{\alpha(0)}{8\pi h^4},
\end{equation}
which is a corollary of Eq.\,(\ref{CP-rCl}).

\section{Planar systems in cylindrical coordinates}
\label{sec-plane}

In Sec.~\ref{sec-planeR} we discussed planar systems using rectangular
coordinates in considerable detail.
Here we discuss planar systems using cylindrical polar coordinates,
\begin{equation}
(z,\rho,\phi), \qquad
-\infty<z<\infty, \quad 0\leq\rho <\infty, \quad 0\leq\phi <2\pi. 
\end{equation}
Here we go through most of the construction from the last section,
using cylindrical coordinates, to gain insight into how the azimuth
modes contribute in axially symmetric systems.
The construction of vector eigenfunctions using cylindrical coordinates
is significantly different and is now described by the azimuth mode $m$.
Developments in the last section related to 
separation of modes, derivation of boundary conditions
and the related Green's dyadic, 
broadly follows the same steps and we shall suffice our discussion
by giving a transcription to go from rectangular coordinates to 
cylindrical coordinates. 

The gradient operator in cylindrical coordinates is given by
\begin{equation}
{\bm\nabla} = \hat{\bf z} \frac{\partial}{\partial z}
+ \hat{\bm\rho} \frac{\partial}{\partial \rho}
+ \hat{\bm\phi} \frac{1}{\rho}\frac{\partial}{\partial \phi}.
\end{equation}
After choosing $\hat{\bf z}$ to be the direction of normal to 
the plane, the following operators yield
two orthogonal vectors perpendicular to $\hat{\bf z}$, 
\begin{subequations}
\begin{align}
{\bm\nabla}_\perp &= \hat{\bm\rho} \frac{\partial}{\partial \rho}
+ \hat{\bm\phi} \frac{1}{\rho}\frac{\partial}{\partial \phi}, \\
\hat{\bf z} \times {\bm\nabla}_\perp 
&= - \hat{\bm\rho} \frac{1}{\rho}\frac{\partial}{\partial \phi}
+ \hat{\bm\phi} \frac{\partial}{\partial \rho}.
\end{align}
\end{subequations}
An arbitrary vector field ${\bf E}({\bf r})$ can be expressed in the form 
of Eq.\,(\ref{ve=thE}), now with the respective gradient operators
in cylindrical coordinates,
where $E^{(i)}({\bf r})$, $i=u,v,w,$ are arbitrary functions. 
For systems with rotational symmetry about the normal direction $\hat{\bf z}$
it is convenient to express the functions $E^{(i)}({\bf r})$
in terms of the eigenfunctions of the Laplacian on the surface of the
planar interface, now in cylindrical coordinates. This Laplacian satisfies
\begin{align}
\nabla_\perp^2 \Big[ e^{im\phi} J_m(k_\perp\rho) \Big]
&= \left( \frac{1}{\rho} \frac{\partial}{\partial \rho}
\rho \frac{\partial}{\partial \rho}
+ \frac{1}{\rho^2} \frac{\partial^2}{\partial \phi^2} \right)
\Big[ e^{im\phi} J_m(k_\perp\rho) \Big]
\nonumber \\
&= e^{im\phi} \left( \frac{1}{\rho} \frac{\partial}{\partial \rho}
\rho \frac{\partial}{\partial \rho}
- \frac{m^2}{\rho^2} \right) J_m(k_\perp\rho) \nonumber \\
&= -k_\perp^2 \Big[ e^{im\phi} J_m(k_\perp\rho) \Big],
\label{ev-pcc}
\end{align}
where we used the differential equation satisfied by the Bessel functions,
\begin{equation}
\left( \frac{1}{\rho} \frac{\partial}{\partial \rho}
\rho \frac{\partial}{\partial \rho}
- \frac{m^2}{\rho^2} + k_\perp^2 \right) J_m(k_\perp\rho) = 0.
\end{equation}
Thus, we learn that trigonometric functions and Bessel functions are
suitable eigenfunctions of the Laplacian on the surface of a plate.
These eigenfunctions, together, satisfy the completeness relation
\begin{equation}
\frac{1}{\rho} \delta(\rho-\rho^\prime) \delta(\phi-\phi^\prime)
= \frac{1}{2\pi} \sum_{m=-\infty}^\infty \int_0^\infty k_\perp dk_\perp 
\,e^{-im(\phi-\phi^\prime)} J_m(k_\perp\rho) J_m(k_\perp\rho^\prime)
\end{equation}
and the orthogonality relation
\begin{equation}
\int_0^{2\pi} d\phi \int_0^\infty \rho d\rho
\,e^{-i\phi(m-m^\prime)} J_m(k_\perp\rho) J_{m^\prime}(k_\perp^\prime\rho)
= 2\pi\delta_{mm^\prime} \frac{\delta(k_\perp-k_\perp^\prime)}{k_\perp}.
\end{equation}
This implies that the functions $E^{(i)}({\bf r})$ can be 
expressed in the form
\begin{equation}
E^{(i)}({\bf r}) = \frac{1}{2\pi} \sum_{m=-\infty}^\infty e^{im\phi}
\int_0^\infty k_\perp dk_\perp J_m(k_\perp\rho) E^{(i)}_m(z,k_\perp),
\label{FouCyl}%
\end{equation}
where the components are given by
\begin{equation}
E^{(i)}_m(z,k_\perp) = \int_0^{2\pi}d\phi\,e^{-im\phi}
\int_0^\infty \rho d\rho \,J_m(k_\perp\rho) \,E^{(i)}({\bf r}).
\end{equation}

\subsection{Vector eigenfunctions}

Using Eqs.~(\ref{FouCyl}) in Eq.\,(\ref{ve=thE}) we can
express the vector field ${\bf E}({\bf r})$ as
\begin{align}
{\bf E}({\bf r}) &= \frac{1}{2\pi} \sum_{m=-\infty}^\infty 
\int_0^\infty k_\perp dk_\perp \bigg[
{\bf U}_m(k_\perp\rho,\phi) E^{(u)}_m(z,k_\perp) 
\nonumber \\ & \hspace{5mm}
+{\bf V}_m(k_\perp\rho,\phi) E^{(v)}_m(z,k_\perp)
+{\bf W}_m(k_\perp\rho,\phi) E^{(w)}_m(z,k_\perp) \bigg],
\label{arV}
\end{align} 
where the cylindrical vector eigenfunctions are defined as
\begin{subequations}
\begin{align}
{\bf U}_m(k_\perp\rho,\phi) &= \frac{1}{ik_\perp}{\bm\nabla}_\perp 
e^{im\phi} \,J_m(k_\perp\rho), \label{Uef} \\
{\bf V}_m(k_\perp\rho,\phi) &= \frac{1}{ik_\perp} 
(\hat{\bf z} \times {\bm\nabla}_\perp)
e^{im\phi} \,J_m(k_\perp\rho), \label{Vef} \\
{\bf W}_m(k_\perp\rho,\phi) &= \hat{\bf z} \, 
e^{im\phi} \,J_m(k_\perp\rho).
\end{align}%
\label{vecEF-p}%
\end{subequations}
The functions $E^{(u)}_m(z,k_\perp)$ and $E^{(v)}_m(z,k_\perp)$ in
Eq.\,(\ref{arV}) are redefinitions of the ones in Eqs.~(\ref{FouCyl})
to accommodate factors of $ik_\perp$. 
We note that the mode corresponding to $k_\perp=0$ and $m=0$ does not 
exist in ${\bf U}_m(k_\perp\rho,\phi)$ and ${\bf V}_m(k_\perp\rho,\phi)$,
for the same reason mentioned after Eqs.\,(\ref{Fvef}).
Thus, we define the vector eigenfunctions
\begin{subequations}
\begin{align}
{\bf X}_m^{(u)}(k_\perp\rho,\phi) &= {\bf U}_m(k_\perp\rho,\phi), \\
\quad {\bf X}_m^{(v)}(k_\perp\rho,\phi) &= {\bf V}_m(k_\perp\rho,\phi), \\
\quad {\bf X}_m^{(w)}(k_\perp\rho,\phi) &= {\bf W}_m(k_\perp\rho,\phi).
\end{align}%
\label{vecEF-pX}%
\end{subequations}%
The vector field of Eq.\,(\ref{arV}) can be written 
in the form
\begin{equation}
{\bf E}({\bf r}) = \frac{1}{2\pi} \sum_{m=-\infty}^\infty
\int_0^\infty k_\perp dk_\perp
{\bf X}_m^{(i)}(k_\perp\rho,\phi) E_m^{(i)}(z,k_\perp).
\end{equation}
The vector eigenfunctions satisfy the completeness relation
\begin{equation}
\frac{1}{\rho}\delta(\phi-\phi^\prime) \delta(\rho - \rho^\prime) {\bf 1}
= \frac{1}{2\pi} \sum_{m=-\infty}^\infty \int_0^\infty k_\perp dk_\perp \,
{\bf X}_m^{(i)}(k_\perp\rho,\phi)
{\bf X}_{m}^{(i)*}(k_\perp\rho^\prime,\phi^\prime)
\label{Com-Vcy}
\end{equation}
and the orthogonality relation
\begin{equation}
\int_0^{2\pi}d\phi \int_0^\infty\rho d\rho \,
{\bf X}_m^{(i)*}(k_\perp\rho,\phi)
\cdot {\bf X}_{m^\prime}^{(j)}(k_\perp^\prime\rho,\phi) = 2\pi \delta^{ij}
\delta_{mm^\prime} \frac{1}{k_\perp} \delta(k_\perp - k_\perp^\prime).
\end{equation}
The unit dyadic appearing in Eq.\,(\ref{Com-Vcy}) is
\begin{equation}
{\bf 1} = {\bm\nabla} {\bf r} = \hat{\bf z} \hat{\bf z}
+ \hat{\bm\rho} \hat{\bm\rho} + \hat{\bm\phi} \hat{\bm\phi}.
\end{equation}

The divergence of a vector field in the basis of
the vector eigenfunctions of Eq.\,(\ref{vecEF-p}) is evaluated as
\begin{equation}
{\bm\nabla}\cdot {\bf E} = \frac{1}{2\pi} \sum_{m=-\infty}^\infty
\int_0^\infty k_\perp dk_\perp e^{im\phi} J_m(k_\perp\rho)
\left[ ik_\perp E^{(u)}_m(z,k_\perp) 
+ \frac{\partial}{\partial z} E^{(w)}_m(z,k_\perp) \right],
\end{equation}
which uses the identities,
\begin{subequations}
\begin{align}
\hat{\bf z} \cdot {\bf U}_m &=0, & {\bm\nabla}_\perp \cdot {\bf U}_m &=
ik_\perp e^{im\phi} J_m(k_\perp\rho), \\ 
\hat{\bf z} \cdot {\bf V}_m &=0, & {\bm\nabla}_\perp \cdot {\bf V}_m &=0, \\
\hat{\bf z} \cdot {\bf W}_m &=
e^{im\phi} J_m(k_\perp\rho), & 
{\bm\nabla}_\perp \cdot {\bf W}_m &=0,
\end{align} 
\end{subequations}
and
\begin{subequations}
\begin{align}
{\bm\nabla} \cdot {\bf U}_m &=
ik_\perp e^{im\phi} J_m(k_\perp\rho), \\ 
{\bm\nabla} \cdot {\bf V}_m &=0, \\ {\bm\nabla} \cdot {\bf W}_m &=
e^{im\phi} J_m(k_\perp\rho) \frac{\partial}{\partial z}.
\end{align}
\end{subequations}
Similarly, the curl of a vector field in the basis of the 
vector eigenfunctions of Eq.\,(\ref{vecEF-p}) is evaluated as
\begin{align}
{\bm\nabla}\times {\bf E} &= \frac{1}{2\pi} \sum_{m=-\infty}^\infty
\int_0^\infty k_\perp dk_\perp \bigg[
- {\bf U}_m(k_\perp\rho,\phi) \frac{\partial}{\partial z} E^{(v)}_m(z,k_\perp)
\nonumber \\ & \hspace{5mm}
+ {\bf V}_m(k_\perp\rho,\phi) 
\left\{ \frac{\partial}{\partial z} E^{(u)}_m(z,k_\perp)
-ik_\perp E^{(w)}_m(z,k_\perp) \right\}
\nonumber \\ & \hspace{5mm}
+ {\bf W}_m(k_\perp\rho,\phi) \,ik_\perp E^{(v)}_m(z,k_\perp) \bigg]
\label{Cu-pl}
\end{align}
using
\begin{subequations}
\begin{align}
\hat{\bf z} \times {\bf U}_m &= {\bf V}_m, &
{\bm\nabla}_\perp \times {\bf U}_m &= 0, \\
\hat{\bf z} \times {\bf V}_m &= -{\bf U}_m, &
{\bm\nabla}_\perp \times {\bf V}_m &= ik_\perp {\bf W}_m, \\ 
\hat{\bf z} \times {\bf W}_m &=0, &
{\bm\nabla}_\perp \times {\bf W}_m &= - ik_\perp {\bf V}_m,
\end{align}
\end{subequations}
and
\begin{subequations}
\begin{align}
{\bm\nabla} \times {\bf U}_m &= {\bf V}_m \frac{\partial}{\partial z}, \\
{\bm\nabla} \times {\bf V}_m &= ik_\perp {\bf W}_m
- {\bf U}_m \frac{\partial}{\partial z}, \\ 
{\bm\nabla} \times {\bf W}_m &= - ik_\perp {\bf V}_m.
\end{align}
\end{subequations}

Using the form for the material properties in Eqs.\,(\ref{emRsym})
and the decomposition of a vector field in terms of the Fourier vector
eigenfunctions in Eq.\,(\ref{Ereck}), together, in Eq.\,(\ref{DB=emuEB})
we obtain the following decomposition of the macroscopic fields
in terms of the Fourier vector eigenfunctions,
\begin{subequations}
\begin{align}
{\bf D}({\bf r}) &= 
\frac{1}{2\pi} \sum_{m=-\infty}^\infty
\int_0^\infty k_\perp dk_\perp \bigg[
{\bf U}_m \varepsilon^\perp(z;\omega) E^{(u)}_m(z,k_\perp)
\nonumber \\ & \hspace{5mm}
+{\bf V}_m \varepsilon^\perp(z;\omega) E^{(v)}_m(z,k_\perp)
+{\bf W}_m \varepsilon^{||}(z;\omega) E^{(w)}_m(z,k_\perp) \bigg], \\
{\bf B}({\bf r}) &= \frac{1}{2\pi} \sum_{m=-\infty}^\infty
\int_0^\infty k_\perp dk_\perp \bigg[
{\bf U}_m \mu^\perp(z;\omega) H^{(u)}_m(z,k_\perp)
\nonumber \\ & \hspace{5mm}
+{\bf V}_m \mu^\perp(z;\omega) H^{(v)}_m(z,k_\perp)
+{\bf W}_m \mu^{||}(z;\omega) H^{(w)}_m(z,k_\perp) \bigg],
\end{align}%
\label{D=eE-pl}%
\end{subequations}%
where we have suppressed the arguments in the vector eigenfunctions 
${\bf U}_m(k_\perp\rho,\phi)$, ${\bf V}_m(k_\perp\rho,\phi)$, 
and ${\bf W}_m(k_\perp\rho,\phi)$, to save typographic space.

\subsection{Separation of modes}

We derived the curl of a vector field in Eq.\,(\ref{Cu-plR}) in the basis 
of rectangular vector eigenfunctions of Eq.\,(\ref{FvefX}),
and again in Eq.\,(\ref{Cu-pl}) in the basis of cylindrical 
vector eigenfunctions of Eq.\,(\ref{vecEF-pX}).
Comparing the two expressions for the curl in Eqs.\,(\ref{Cu-plR}) and 
(\ref{Cu-pl}) it is remarkable that both have the same structure, 
and the differences are completely contained inside the vector 
eigenfunctions. The transcription to switch from rectangular 
to cylindrical coordinates is
\begin{subequations}
\begin{align}
{\bf X}^{(i)}({\bf k}_\perp,{\bf r}_\perp)
&\to {\bf X}^{(i)}_m(k_\perp\rho,\phi), \\
E^{(i)}(z,{\bf k}_\perp) &\to E^{(i)}_m(z,k_\perp), \\
\int \frac{d^2k_\perp}{(2\pi)^2} &\to
\frac{1}{2\pi} \sum_{m=-\infty}^\infty \int_0^\infty k_\perp dk_\perp.
\end{align}%
\label{trn-re-cy}%
\end{subequations}%
This feature is not merely a consequence of the system of planar
geometry being expressed in two different coordinate systems. In 
Sec.~\ref{sec-sphere} we shall show that for spherical vector eigenfunctions 
the curl operator again has a similar structure. Since the Maxwell
equations in the absence of free charges and currents are completely 
given by the curl of the fields, we conclude that a decomposition 
of the curl of a vector field in the particular form of Eq.\,(\ref{Cu-plR}),
in a particular basis of vector eigenfunctions, guarantees the separation 
of electromagnetic modes into transverse electric and magnetic modes.

Thus, all the developments leading to the separation of modes
in Eqs.(\ref{ME1-plR}) to (\ref{bc-plR}),
and the construction of the Green's dyadic 
in Eqs.\,(\ref{EM-Gfun}) to (\ref{Phi=gH}), goes through identically,
with the transcription in Eq.\,(\ref{trn-re-cy}).
We observe that because the Green functions 
$g^{E,H}(z,z^\prime;k_\perp,\omega)$ and the associated Green dyadics
$\gamma^{ij}(z,z^\prime;k_\perp,\omega)$ and
$\phi^{ij}(z,z^\prime;k_\perp,\omega)$ have dependence of the form
$k_\perp =|{\bf k}_\perp|$, the corresponding functions and dyadics
in cylindrical coordinates have no $m$ dependence. That is, only
the $m=0$ mode contributes. 

\subsection{Casimir-Polder interaction energy}

The interaction energy between a polarizable molecule of
Eq.\,(\ref{pol-atom-e1}) and a $\delta$-function dielectric plate
of Eq.\,(\ref{dp-pol-def}), illustrated in Fig.~\ref{fig-dplate-atom-CP12},
is given by, using the transcription 
in Eq.\,(\ref{trn-re-cy}) in Eq.\,(\ref{cp-mcde-plR}),
\begin{equation}
E = - \int_{-\infty}^\infty d\zeta \int_0^\infty k_\perp dk_\perp \,
\gamma^{ii^\prime}(h,a;k_\perp,\omega) \lambda^{i^\prime j^\prime}(\omega)
\gamma_{0}^{j^\prime j}(a,h;k_\perp,\omega) \alpha^{ji}(\omega),
\label{Ecp=cyl12}
\end{equation}
where the dependence in $m$ is completely inside $\alpha^{ji}$,
given in terms of the cylindrical vector eigenfunctions,
\begin{align}
\alpha^{ji}(\omega) &= \frac{1}{2\pi} \sum_{m=-\infty}^\infty
{\bf X}_m^{(j)*}(k_\perp\rho_0,\phi_0)
\cdot {\bm\alpha}(\omega) \cdot {\bf X}_m^{(i)}(k_\perp\rho_0,\phi_0)
\nonumber \\ &
= \frac{1}{2\pi} \sum_{m=-\infty}^\infty \left[ \begin{array}{ccc}
{\bf U}_m^* \cdot {\bm\alpha}(\omega) \cdot {\bf U}_m &
{\bf U}_m^* \cdot {\bm\alpha}(\omega) \cdot {\bf V}_m &
{\bf U}_m^* \cdot {\bm\alpha}(\omega) \cdot {\bf W}_m \\
{\bf V}_m^* \cdot {\bm\alpha}(\omega) \cdot {\bf U}_m &
{\bf V}_m^* \cdot {\bm\alpha}(\omega) \cdot {\bf V}_m &
{\bf V}_m^* \cdot {\bm\alpha}(\omega) \cdot {\bf W}_m \\
{\bf W}_m^* \cdot {\bm\alpha}(\omega) \cdot {\bf U}_m &
{\bf W}_m^* \cdot {\bm\alpha}(\omega) \cdot {\bf V}_m &
{\bf W}_m^* \cdot {\bm\alpha}(\omega) \cdot {\bf W}_m \\
\end{array} \right],
\label{aji-gend3}
\end{align}
where the arguments of 
${\bf U}_m(k_\perp\rho_0,\phi_0)$, ${\bf V}_m(k_\perp\rho_0,\phi_0)$, and
${\bf W}_m(k_\perp\rho_0,\phi_0)$ has been suppressed.

We can use the addition theorem for Bessel functions,
\begin{equation}
\sum_{m=-\infty}^\infty e^{im(\phi-\phi^\prime)}
J_m(k_\perp\rho)J_m(k_\perp\rho^\prime) = J_0(k_\perp P),
\end{equation}
where $P=|{\bf r}_\perp-{\bf r}_\perp^\prime|$, to derive
the following addition theorems for the vector eigenfunctions,
\begin{align}
\frac{1}{2\pi} &= \sum_{m=-\infty}^\infty
{\bf U}_m^*(k_\perp\rho,\phi) \cdot {\bf U}_m(k_\perp\rho,\phi)
\nonumber \\ &= \sum_{m=-\infty}^\infty
{\bf V}_m^*(k_\perp\rho,\phi) \cdot {\bf V}_m(k_\perp\rho,\phi)
\nonumber \\ &= \sum_{m=-\infty}^\infty
{\bf W}_m^*(k_\perp\rho,\phi) \cdot {\bf W}_m(k_\perp\rho,\phi).
\label{AddTh-VE-pl}%
\end{align}
For an isotropically polarizable molecule we have 
${\bm\alpha}(\omega) ={\bf 1}\alpha(\omega)$. For this scenario,
using the addition theorems in Eq.~(\ref{AddTh-VE-pl}),
and the orthogonality feature of the vector eigenfunctions when they
are evaluated at the same position, based on Eq.\,(\ref{gr-id-fg}),
we have
\begin{equation}
\alpha^{ji}(\omega) = \delta^{ji} \frac{\alpha(\omega)}{2\pi},
\end{equation}
which leads to the Casimir-Polder result~\cite{Casmir:1947hx} for
isotropically polarizable molecule, 
\begin{equation}
E=-\frac{3\alpha(0)}{8\pi h^4},
\end{equation}
a corollary of Eq.\,(\ref{CP-rCl}).

Without any loss of generality we can choose the position of the
polarizable molecule to be on the $z$ axis, 
such that we have $\rho=0$ and $\phi=0$. For this case,
in terms of a unit vector $\hat{\bf t}$ evaluated on the axis,
\begin{equation}
\hat{\bf t} = \frac{1}{\sqrt{2}} (\hat{\bm\rho} + i\hat{\bm\phi}),
\end{equation}
we have
\begin{subequations}
\begin{align}
{\bf U}_m(0,0) &= -\dfrac{i}{\sqrt{4\pi}} 
\Big[ \delta_{m,1} \,\hat{\bf t} -\delta_{m,-1} \,\hat{\bf t}^* \Big], \\
{\bf V}_m(0,0) &= -\dfrac{1}{\sqrt{4\pi}} 
\Big[ \delta_{m,1} \,\hat{\bf t} +\delta_{m,-1} \,\hat{\bf t}^* \Big],
\end{align}%
\label{uvm=0012}
\end{subequations}%
and
\begin{equation}
{\bf W}_m(0,0) =
\dfrac{\delta_{m0}}{\sqrt{2\pi}} \,\hat{\bf z},
\end{equation}
using 
\begin{subequations}
\begin{align}
J_m^\prime(0) &= \dfrac{1}{2}
\Big[ \delta_{m,1} -\delta_{m,-1} \Big], \\
\lim_{x\to 0} \frac{m J_m(x)}{x} &= \dfrac{1}{2}
\Big[ \delta_{m,1} +\delta_{m,-1} \Big].
\end{align}
\end{subequations}
Using Eqs.\,(\ref{uvm=0012}) we can evaluate
\begin{align}
\alpha^{ji}(\omega) &= \frac{1}{2\pi} \sum_{m=-\infty}^\infty
{\bf X}_m^{(j)*}(0,0) \cdot {\bm\alpha}(\omega) \cdot {\bf X}_m^{(i)}(0,0)
\nonumber \\ &
= \frac{1}{2\pi} \left[ \begin{array}{ccc}
\frac{1}{2} \text{tr}\left\{{\bm\alpha}(\omega) \cdot {\bf 1}_\perp\right\} 
& 0 & 0 \\ 0 & 
\frac{1}{2} \text{tr}\left\{{\bm\alpha}(\omega) \cdot {\bf 1}_\perp\right\} 
& 0 \\ 0 & 0 &
\text{tr}\left\{{\bm\alpha}(\omega) \cdot \hat{\bf z} \hat{\bf z} \right\}
\end{array} \right].
\label{af-axis}
\end{align}
Using Eq.\,(\ref{af-axis}) in Eq.\,(\ref{Ecp=cyl12}) 
we reproduce the result for the energy in Eq.\,(\ref{CP-pl-genEan})
between a polarizable molecule and $\delta$-function plate.
All the related results associated with Eq.\,(\ref{CP-pl-genEan}) follow.

Thus, we have derived the Casimir-Polder energy of Eq.\,(\ref{CP-pl-genEan})
using cylindrical coordinates and the respective
cylindrical vector eigenfunctions.
The planar case, of course, is simple, and it is not surprising that all
the $m$ modes separate in this case. But, it has given us insight into
how the $m=0$ mode alone contributes to the Casimir-Polder energy. 


\section{Spherically symmetric systems}
\label{sec-sphere}

\begin{figure}%
\begin{center}
\includegraphics{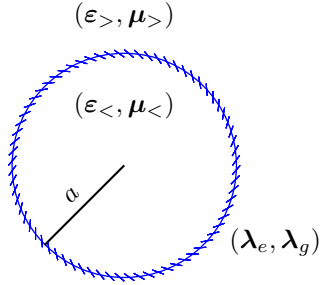}
\caption{A spherically symmetric $\delta$-function sphere
described by properties $({\bm\lambda}_e,{\bm\lambda}_g)$
sandwiched at the interface of a spherically symmetric dielectric ball
of properties $({\bm\varepsilon}_<,{\bm\mu}_<)$
immersed in another spherically symmetric medium of infinite extent
with properties $({\bm\varepsilon}_>,{\bm\mu}_>)$.
The material properties described by the dielectric permittivity
and magnetic permeability are given by Eqs.~(\ref{epmu-def-sp}).}%
\label{Dielectric-ball-fig}.%
\end{center}%
\end{figure}

We next study systems with spherical symmetry.
For example, here, we will consider the system described in 
Fig.~\ref{Dielectric-ball-fig}.
The following study closely follows our study of systems with planar
symmetry in Sec.~\ref{sec-plane}.
We shall notice that it is possible to transcribe from the equations
for planar symmetry to the corresponding equations for spherical symmetry.
But, for pedagogical purpose, at the cost of plausible repetition of 
the concepts, we illustrate the procedure for the the case of 
spherically symmetric systems. 

\subsection{Spherical vector eigenfunctions}

We shall use the spherical polar coordinates,
\begin{equation}
(r,\theta,\phi), \qquad 0\leq r<\infty, \quad
0\leq \theta \leq\pi, \quad 0\leq \theta <2\pi.
\end{equation}
The gradient operator in spherical coordinates is
\begin{equation}
{\bm\nabla} = \hat{\bf r} \frac{\partial}{\partial r}
+ \hat{\bm\theta} \frac{1}{r} \frac{\partial}{\partial \theta}
+ \hat{\bm\phi} \frac{1}{r\sin\theta}\frac{\partial}{\partial \phi}.
\end{equation}
We can construct the operators
\begin{subequations}
\begin{align}
{\bm\nabla}_\perp &=
\hat{\bm\theta} \frac{1}{r} \frac{\partial}{\partial \theta}
+ \hat{\bm\phi} \frac{1}{r\sin\theta}\frac{\partial}{\partial \phi}, \\
\hat{\bf r} \times {\bm\nabla}_\perp 
&= - \hat{\bm\theta} \frac{1}{r\sin\theta}\frac{\partial}{\partial \phi}
+ \hat{\bm\phi} \frac{1}{r} \frac{\partial}{\partial \theta},
\end{align}
\end{subequations}
which together with the radial direction $\hat{\bf r}$
forms an orthogonal set of vectors.
An arbitrary vector field can thus be expressed in the form
\begin{equation}
{\bf E}({\bf r}) = {\bm\nabla}_\perp E^{(u)}({\bf r})
+ \hat{\bf r}\times {\bm\nabla}_\perp E^{(v)}({\bf r})
+ \hat{\bf r} \,E^{(w)}({\bf r}).
\label{Vec-sp}
\end{equation}
The Laplacian on the surface of a sphere, 
transverse to $\hat{\bf r}$, is given by 
\begin{equation}
\nabla_\perp^2 = \frac{1}{r^2} \left(
\frac{1}{\sin\theta} \frac{\partial}{\partial \theta}
\sin\theta \frac{\partial}{\partial \theta}
+ \frac{1}{\sin^2\theta} \frac{\partial^2}{\partial \phi^2} \right).
\end{equation}
It satisfies the eigenvalue equation
\begin{equation}
\nabla_\perp^2 Y_{lm}(\theta,\phi) 
= - \frac{l(l+1)}{r^2} Y_{lm}(\theta,\phi), 
\label{ev-scc}
\end{equation}
where the spherical harmonics $Y_{lm}(\theta,\phi)$ are defined as
\begin{equation}
Y_{lm}(\theta,\phi) = \frac{e^{im\phi}}{\sqrt{2\pi}} \,\Theta_{lm}(\theta), 
\end{equation}
where
\begin{equation}
\Theta_{lm}(\theta) = \sqrt{\frac{2l+1}{2}} \sqrt{\frac{(l+m)!}{(l-m)!}}
\frac{1}{(\sin\theta)^m} \left[ \frac{d}{d\cos\theta} \right]^{l-m}
\frac{(\cos^2\theta -1)^l}{2^ll!}.
\end{equation}
The spherical harmonics satisfy the completeness relation
\begin{equation}
\frac{1}{\sin\theta} \delta(\theta-\theta^\prime) \delta(\phi-\phi^\prime)
= \sum_{l=0}^\infty \sum_{m=-l}^l 
Y_{lm}(\theta,\phi) Y_{lm}^*(\theta^\prime,\phi^\prime)
\end{equation}
and the orthonormality relation
\begin{equation}
\int_0^{2\pi}d\phi \int_0^{\pi}\sin\theta d\theta\,
Y_{lm}^*(\theta,\phi) Y_{l^\prime m^\prime}(\theta,\phi)
= \delta_{ll^\prime} \delta_{mm^\prime},
\end{equation}
which together implies that any function on the surface of a sphere can be 
expanded in terms of the spherical harmonics.
Thus, we can express the functions $E^{(i)}({\bf r})$ appearing
in Eq.\,(\ref{Vec-sp}) in terms of the spherical harmonics as
\begin{subequations}
\begin{align}
E^{(i)}({\bf r}) &= \sum_{l=0}^\infty \sum_{m=-l}^l 
Y_{lm}(\theta,\phi) E^{(i)}_{lm}(r), \\
E^{(i)}_{lm}(r) &= \int_0^{2\pi}d\phi \int_0^{\pi}\sin\theta d\theta
\,Y_{lm}^*(\theta,\phi) \,E^{(i)}({\bf r}),
\end{align}%
\label{FouSp}%
\end{subequations}
where $i=u,v,w$. Using Eqs.~(\ref{FouSp}) in Eq.\,(\ref{Vec-sp}) we can
express the vector field ${\bf E}({\bf r})$ as
\begin{equation}
{\bf E}({\bf r}) = \sum_{l=0/1}^\infty \sum_{m=-l}^l \bigg[
{\bf U}_{lm}(\theta,\phi) E^{(u)}_{lm}(r)
+{\bf V}_{lm}(\theta,\phi) E^{(v)}_{lm}(r)
+{\bf W}_{lm}(\theta,\phi) E^{(w)}_{lm}(r) \bigg],
\label{arVsp}
\end{equation} 
where the spherical vector eigenfunctions are defined as
\begin{subequations}
\begin{align}
{\bf U}_{lm}(\theta,\phi) &= \frac{1}{ik_\perp}{\bm\nabla}_\perp 
Y_{lm}(\theta,\phi), \label{Uefsp} \\
{\bf V}_{lm}(\theta,\phi) &= \frac{1}{ik_\perp} 
\hat{\bf r} \times {\bm\nabla}_\perp
Y_{lm}(\theta,\phi), \label{Vefsp} \\
{\bf W}_{lm}(\theta,\phi) &= \hat{\bf r} \, 
Y_{lm}(\theta,\phi). \label{Wefsp}
\end{align}%
\label{vec-EF-sp}%
\end{subequations}
The $ik_\perp$'s in the denominators of the vector eigenfunctions
in Eqs.~(\ref{Uefsp}) and (\ref{Vefsp}) are defined as
\begin{equation}
k_\perp^2 = \frac{l(l+1)}{r^2},
\label{kperp-def}
\end{equation}
which are introduced to normalize the vector eigenfunctions. 
This normalization seems to involve division by zero for $l=0$.
But, the mode corresponding to $l=0$ and $m=0$ does not exist in
the constructions of ${\bf U}_{lm}$ and ${\bf V}_{lm}$, because
the gradient operator gives zero in their operations for this mode.
Unlike the similar situations encountered in the rectangular and cylindrical
vector eigenfunctions, refer discussions following Eqs.\,(\ref{Fvef}),
we have to be careful about this fact for spherical vector eigenfunctions.
That is, the sum over $l$ runs from $0$ to $\infty$ for the terms involving
${\bf W}_{lm}$, but $l$ runs from $1$ to $\infty$ for terms involving 
${\bf U}_{lm}$ and ${\bf V}_{lm}$. We emphasize on this by explicitly 
writing $l=0/1$ for the initial value of index $l$ in the summation 
in Eq.\,(\ref{arVsp}) and in what follows. 
The functions $E^{(u)}_{lm}(z,k_\perp)$ and $E^{(v)}_{lm}(z,k_\perp)$ in
Eq.\,(\ref{arVsp}), relative to the ones in Eqs.~(\ref{FouSp}),
are suitably redefined to accommodate this factor of $ik_\perp$.
Then, using the notations,
\begin{subequations}
\begin{align}
{\bf X}_{lm}^{(u)}(\theta,\phi) &= {\bf U}_{lm}(\theta,\phi), \\
\quad {\bf X}_{lm}^{(v)}(\theta,\phi) &= {\bf V}_{lm}(\theta,\phi), \\
\quad {\bf X}_{lm}^{(w)}(\theta,\phi) &= {\bf W}_{lm}(\theta,\phi),
\end{align}
\end{subequations}
the vector field ${\bf E}({\bf r})$ of Eq.\,(\ref{arVsp}) can be written
in the form
\begin{equation}
{\bf E}({\bf r}) = \sum_{l=0/1}^\infty \sum_{m=-l}^l
{\bf X}_{lm}^{(i)}(\theta,\phi) E_{lm}^{(i)}(r),
\end{equation}
where summation convention in the index $i$ is assumed.
The spherical vector eigenfunctions of Eq.\,(\ref{vec-EF-sp})
satisfy the completeness relation
\begin{equation}
\frac{1}{\sin\theta}\delta(\theta-\theta^\prime) \delta(\phi-\phi^\prime)
{\bf 1} = \sum_{l=0/1}^\infty \sum_{m=-l}^l \,
{\bf X}_{lm}^{(i)}(\theta,\phi)
{\bf X}_{lm}^{*(i)}(\theta^\prime,\phi^\prime)
\label{Com-Vsp}
\end{equation}
and the orthonormality relation
\begin{equation}
\int_0^{2\pi}d\phi \int_0^{\pi}\sin\theta d\theta \,
{\bf X}_{lm}^{(i)}(\theta,\phi)
\cdot {\bf X}_{l^\prime m^\prime}^{(j)}(\theta,\phi) = \delta^{ij}
\delta_{ll^\prime} \delta_{mm^\prime},
\end{equation}
where the unit dyadic appearing in Eq.\,(\ref{Com-Vsp}) is
\begin{equation}
{\bf 1} = {\bm\nabla} {\bf r} = \hat{\bf r} \hat{\bf r}
+ \hat{\bm\theta} \hat{\bm\theta} + \hat{\bm\phi} \hat{\bm\phi}.
\end{equation}

The divergence of a vector field 
in the basis of spherical vector eigenfunctions of Eq.\,(\ref{vec-EF-sp})
is evaluated as
\begin{equation}
{\bm\nabla}\cdot {\bf E} = \sum_{l=0}^\infty \sum_{m=-l}^l \, 
Y_{lm}(\theta,\phi) \left[ ik_\perp E^{(u)}_{lm}(r) 
+\left(\frac{2}{r} +\frac{\partial}{\partial r}\right) E^{(w)}_{lm}(r) \right],
\end{equation}
which uses the identities,
\begin{subequations}
\begin{align}
\hat{\bf r} \cdot {\bf U}_{lm} &=0, 
& {\bm\nabla}_\perp \cdot {\bf U}_{lm} &= ik_\perp Y_{lm}(\theta,\phi), \\ 
\hat{\bf r} \cdot {\bf V}_{lm} &=0, & 
{\bm\nabla}_\perp \cdot {\bf V}_{lm} &=0, \\ 
\hat{\bf r} \cdot {\bf W}_{lm} &= Y_{lm}(\theta,\phi), &
{\bm\nabla}_\perp \cdot {\bf W}_{lm} &= \frac{2}{r} Y_{lm}(\theta,\phi),
\end{align}
\end{subequations}
and
\begin{subequations}
\begin{align}
{\bm\nabla} \cdot {\bf U}_{lm} &= ik_\perp Y_{lm}(\theta,\phi), \\ 
{\bm\nabla} \cdot {\bf V}_{lm} &=0, \\ 
{\bm\nabla} \cdot {\bf W}_{lm} &= Y_{lm}(\theta,\phi) 
\left( \frac{2}{r} + \frac{\partial}{\partial r} \right).
\end{align}
\end{subequations}
Similarly, the curl of a vector field
in the basis of spherical vector eigenfunctions of Eq.\,(\ref{vec-EF-sp})
is evaluated as
\begin{align}
{\bm\nabla}\times {\bf E} = \sum_{l=0/1}^\infty \sum_{m=-l}^l \, \bigg[ &
- {\bf U}_{lm}(\theta,\phi) 
\left( \frac{1}{r} + \frac{\partial}{\partial r} \right) E^{(v)}_{lm}(r)
\nonumber \\ &
+ {\bf V}_{lm}(\theta,\phi) 
\left\{ \left( \frac{1}{r} + \frac{\partial}{\partial r} \right)
E^{(u)}_{lm}(r) -ik_\perp E^{(w)}_{lm}(r) \right\}
\nonumber \\ &
+ {\bf W}_{lm}(\theta,\phi) \,ik_\perp E^{(v)}_{lm}(r) \bigg]
\label{Cu-sp}
\end{align}
using
\begin{subequations}
\begin{align}
\hat{\bf r} \times {\bf U}_{lm} &= {\bf V}_{lm}, &
{\bm\nabla}_\perp \times {\bf U}_{lm} &= {\bf V}_{lm} \frac{1}{r}, \\ 
\hat{\bf r} \times {\bf V}_{lm} &= -{\bf U}_{lm}, &
{\bm\nabla}_\perp \times {\bf V}_{lm} 
&= -{\bf U}_{lm} \frac{1}{r} + ik_\perp {\bf W}_{lm}, \\ 
\hat{\bf r} \times {\bf W}_{lm} &=0, & 
{\bm\nabla}_\perp \times {\bf W}_{lm} &= - ik_\perp {\bf V}_{lm},
\end{align}
\end{subequations}
and
\begin{subequations}
\begin{align}
{\bm\nabla} \times {\bf U}_{lm} &= {\bf V}_{lm} 
\left( \frac{1}{r} + \frac{\partial}{\partial r} \right), \\ 
{\bm\nabla} \times {\bf V}_{lm} &= ik_\perp {\bf W}_{lm}
- {\bf U}_{lm} \left( \frac{1}{r} + \frac{\partial}{\partial r} \right), \\ 
{\bm\nabla} \times {\bf W}_{lm} &= - ik_\perp {\bf V}_{lm}.
\end{align}
\end{subequations}
We observe that the curl operator in the basis of spherical 
vector eigenfunctions, given in Eq.\,(\ref{Cu-sp}),
can be obtained from the curl operator in the basis of
planar vector eigenfunctions, given in Eq.\,(\ref{Cu-pl}), by the replacement
\begin{equation}
\frac{\partial}{\partial z} \to
\left( \frac{1}{r} + \frac{\partial}{\partial r} \right)
= \frac{1}{r} \frac{\partial}{\partial r} r
\end{equation}
which can be used to transcribe the equations between the two geometries
because the Maxwell equations of Eqs.~(\ref{ME-cross})
only involves the curl operation.
This correspondence should be noted in light of the observations and
discussions around Eq.\,(\ref{trn-re-cy}). Our notion is that
the the specific form and structure of the curl operation in
Eqs.\,(\ref{Cu-plR}), (\ref{Cu-pl}), and (\ref{Cu-sp}),
is a generic feature that guarantees the separation of electric
and magnetic modes in the Maxwell equations.

\subsection{Separation of modes}

Restricting our study to spherically symmetric materials,
described by the components of dyadics for
dielectric permittivity and magnetic permeability
being diagonal matrices,
\begin{subequations}
\begin{equation}
\varepsilon^{ij}(r;\omega) = \left[ \begin{array}{ccc}
\varepsilon^\perp(r;\omega) & 0 & 0 \\ 0 & \varepsilon^\perp(r;\omega) & 0 \\
0 & 0 & \varepsilon^{||}(r;\omega) \end{array} \right]
\end{equation}
and
\begin{equation}
\mu^{ij}(r;\omega) = \left[ \begin{array}{ccc}
\mu^\perp(r;\omega) & 0 & 0 \\ 0 & \mu^\perp(r;\omega) & 0 \\
0 & 0 & \mu^{||}(r;\omega) \end{array} \right],
\end{equation}
\end{subequations}
leads to the construction of the fields, using Eq.\,(\ref{DB=emuEB}),
\begin{subequations}
\begin{align}
{\bf D}({\bf r}) &= \sum_{l=0/1}^\infty \sum_{m=-l}^l \bigg[
{\bf U}_{lm}(\theta,\phi) \varepsilon^\perp(r;\omega) E^{(u)}_{lm}(r)
\nonumber \\ & \hspace{5mm}
+{\bf V}_{lm}(\theta,\phi) \varepsilon^\perp(r;\omega) E^{(v)}_{lm}(r)
+{\bf W}_{lm}(\theta,\phi) \varepsilon^{||}(r;\omega) E^{(w)}_{lm}(r) \bigg], \\
{\bf B}({\bf r}) &= \sum_{l=0/1}^\infty \sum_{m=-l}^l \bigg[
{\bf U}_{lm}(\theta,\phi) \mu^\perp(r;\omega) H^{(u)}_{lm}(r)
\nonumber \\ & \hspace{5mm}
+{\bf V}_{lm}(\theta,\phi) \mu^\perp(r;\omega) H^{(v)}_{lm}(r)
+{\bf W}_{lm}(\theta,\phi) \mu^{||}(r;\omega) H^{(w)}_{lm}(r) \bigg].
\end{align}%
\label{D=eE-sp}%
\end{subequations}

Using the expression for curl operator in Eq.\,(\ref{Cu-sp}),
in conjunction with Eqs.~(\ref{D=eE-sp}),
the projections of Maxwell's equations in Eqs.~(\ref{ME-cross})
along the spherical vector eigenfunctions, 
${\bf U}_{lm}$, ${\bf V}_{lm}$, and ${\bf W}_{lm}$, are
\begin{subequations}
\begin{align}
&& \underline{{\bm\nabla} \times {\bf E} = i\omega {\bf B}} & \nonumber \\
{\bf U}_{lm} &:& -\frac{1}{r}\frac{\partial}{\partial r}r E^{(v)}_{lm}
&= i\omega\mu^\perp H^{(u)}_{lm}, \label{Max1-U-sp} \\
{\bf V}_{lm} &:& \frac{1}{r}\frac{\partial}{\partial r}r
E^{(u)}_{lm} -ik_\perp E^{(w)}_{lm} &= i\omega\mu^\perp H^{(v)}_{lm},
\label{Max1-V-sp} \\
{\bf W}_{lm} &:& ik_\perp E^{(v)}_{lm} &= i\omega\mu^{||} H^{(w)}_{lm},
\hspace{15mm} \label{Max1-W-sp}
\end{align}
\end{subequations}
and
\begin{subequations}
\begin{align}
&& \underline{{\bm\nabla} \times {\bf H} = -i\omega ({\bf D} + {\bf P})} &
\nonumber \\ 
{\bf U}_{lm} &:& - \frac{1}{r}\frac{\partial}{\partial r}r H^{(v)}_{lm}
&= -i\omega (\varepsilon^\perp E^{(u)}_{lm} + P^{(u)}_{lm}), 
\label{Max2-U-sp} \\
{\bf V}_{lm} &:& 
\frac{1}{r}\frac{\partial}{\partial r}r
H^{(u)}_{lm} -ik_\perp H^{(w)}_{lm} 
&= -i\omega (\varepsilon^\perp E^{(v)}_{lm} +P^{(v)}_{lm}), 
\label{Max2-V-sp} \\
{\bf W}_{lm} &:& ik_\perp H^{(v)}_{lm}
&= -i\omega (\varepsilon^{||} E^{(w)}_{lm} + P^{(w)}_{lm}). \label{Max2-W-sp}
\end{align}
\end{subequations}

\subsection{Boundary conditions on a $\delta$-function sphere}

The boundary conditions satisfied by the fields are derived by integrating
the Maxwell equations across the spherical interface, at $r=a$, to yield 
\cite{Parashar:2017sgo}
\begin{subequations}
\begin{align}
E^{(u)}_{lm}(r) \Big|^{r=a+}_{r=a-}
&= i\omega \lambda_g^\perp H^{(v)}_{lm}(a), \label{bc1-u-sp} \\
E^{(v)}_{lm}(r) \Big|^{r=a+}_{r=a-}
&= -i\omega \lambda_g^\perp H^{(u)}_{lm}(a), \label{bc1-v-sp} \\
D^{(w)}_{lm}(r) \Big|^{r=a+}_{r=a-}
&= -ik_\perp \lambda_e^\perp E^{(u)}_{lm}(a), \label{bc1-w-sp}
\end{align}%
\label{bc1-sp}%
\end{subequations}%
and
\begin{subequations}
\begin{align}
H^{(u)}_{lm}(r) \Big|^{r=a+}_{r=a-} 
&= -i\omega \lambda_e^\perp E^{(v)}_{lm}(a), \label{bc2-u-sp} \\
H^{(v)}_{lm}(r) \Big|^{r=a+}_{r=a-} 
&= i\omega \lambda_e^\perp E^{(u)}_{lm}(a), \label{bc2-v-sp} \\
B^{(w)}_{lm}(r) \Big|^{r=a+}_{r=a-} 
&= -ik_\perp \lambda_g^\perp H^{(u)}_{lm}(a). \label{bc2-w-sp}
\end{align}%
\label{bc2-sp}%
\end{subequations}%
We also obtain the constraints
\begin{subequations}
\begin{align}
\lambda_e^{||} E^{(w)}_{lm}(a) = 0, \label{bc1-cons-sp} \\
\lambda_g^{||} H^{(w)}_{lm}(a) = 0. \label{bc2-cons-sp}
\end{align}%
\label{bc-sp}%
\end{subequations}

\subsection{Spherical Green's dyadic}

In terms of the magnetic and electric spherical Green's functions
defined using
\begin{subequations}
\begin{align}
\left[ -\frac{1}{r} \frac{\partial}{\partial r} r \frac{1}{\mu^\perp(r)}
\frac{1}{r} \frac{\partial}{\partial r} r 
+ \frac{l(l+1)}{r^2\mu^{||}(r)} - \omega^2 \varepsilon^\perp(r)
\right] g_l^E(r,r^\prime;\omega) &= \frac{1}{r^2}\delta(r-r^\prime), \\
\left[ -\frac{1}{r} \frac{\partial}{\partial r} r \frac{1}{\varepsilon^\perp(r)}
\frac{1}{r} \frac{\partial}{\partial r} r 
+ \frac{l(l+1)}{r^2\varepsilon^{||}(r)} - \omega^2 \mu^\perp(r)
\right] g_l^H(r,r^\prime;\omega) &= \frac{1}{r^2}\delta(r-r^\prime),
\label{EM-GfunH-sp}%
\end{align}%
\label{EM-Gfun-sp}%
\end{subequations}
the electric and magnetic fields are expressed in terms of the Green dyadic 
using Eqs.~(\ref{E=GP-def}) and (\ref{H=GP-def}), in the basis of 
the spherical vector eigenfunctions, as
\begin{subequations}
\begin{align}
{\bm\Gamma}({\bf r},{\bf r}^\prime;\omega) &= \sum_{l=0/1}^\infty \sum_{m=-l}^l 
{\bf X}_{lm}^{(i)}(\theta,\phi) \gamma_{lm}^{ij}(r,r^\prime;\omega) 
{\bf X}_{lm}^{(j)*}(\theta^\prime,\phi^\prime), \\
{\bm\Phi}({\bf r},{\bf r}^\prime;\omega) &= \sum_{l=0/1}^\infty \sum_{m=-l}^l 
{\bf X}_{lm}^{(i)}(\theta,\phi) \phi_{lm}^{ij}(r,r^\prime;\omega)
{\bf X}_{lm}^{(j)*}(\theta^\prime,\phi^\prime), 
\end{align}%
\label{G=XgX-sp}%
\end{subequations}
where the matrices $\gamma_{lm}^{ij}(r,r^\prime;\omega)$ and
$\phi_{lm}^{ij}(r,r^\prime;\omega)$ represent the components of the dyadics
in the basis of spherical vector eigenfunctions given by
\begin{subequations}
\begin{align}
\gamma_{lm}^{ij}(r,r^\prime;\omega)
&= \left[ \begin{array}{ccc} \frac{1}{\varepsilon^\perp(r)} 
\frac{1}{r} \frac{\partial}{\partial r} r 
\frac{1}{\varepsilon^\perp(r^\prime)} 
\frac{1}{r^\prime} \frac{\partial}{\partial r^\prime} r^\prime 
 g_l^H & 0 & \frac{1}{\varepsilon^\perp(r)} 
\frac{1}{r} \frac{\partial}{\partial r} r 
\frac{ik_\perp^\prime}{\varepsilon^{||}(r^\prime)} g_l^H \\[2mm]
0 & \omega^2 g_l^E & 0 \\[2mm]
-\frac{ik_\perp}{\varepsilon^{||}(r)} \frac{1}{\varepsilon^\perp(r^\prime)} 
\frac{1}{r^\prime} \frac{\partial}{\partial r^\prime} r^\prime 
g_l^H & 0 & -\frac{ik_\perp}{\varepsilon^{||}(r)}
\frac{ik_\perp^\prime}{\varepsilon^{||}(r^\prime)} g_l^H
\end{array} \right],
\label{Gamma=gE-sp}
\\[3mm]
\phi_{lm}^{ij}(r,r^\prime;\omega)
&= i\omega \left[ \begin{array}{ccc} 0 & \frac{1}{\mu^\perp(r)} 
\frac{1}{r} \frac{\partial}{\partial r} r g_l^E & 0 \\[2mm]
\frac{1}{\varepsilon^\perp(r^\prime)}
\frac{1}{r^\prime} \frac{\partial}{\partial r^\prime} r^\prime 
g_l^H & 0 & \frac{ik_\perp^\prime}{\varepsilon^{||}(r^\prime)} g_l^H \\[2mm]
0 & -\frac{ik_\perp}{\mu^{||}(r)} g_l^E & 0
\end{array} \right],
\label{Phi=gH-sp}
\end{align}
\end{subequations}
with $k_\perp$ defined using Eq.\,(\ref{kperp-def}) and
\begin{equation}
k_\perp^{\prime\, 2} = \frac{l(l+1)}{{r^\prime}^2}.
\label{kperpp-def}
\end{equation}
We have omitted a term containing a $\delta$-function 
in Eq.\,(\ref{Gamma=gE-sp}), which does not contribute to 
interaction energies between disjoint bodies. The explicit form of this
term is
\begin{equation}
- \frac{1}{r^2} \delta(r-r^\prime) \left[ \begin{array}{ccc}
\frac{1}{\varepsilon^\perp(r)} & 0 & 0 \\ 0 & 0 & 0 \\
0 & 0 & \frac{1}{\varepsilon^{||}(r)} \end{array} \right],
\end{equation}
which is presented here for completeness.

\subsection{Magnetic spherical Green's function}
\label{sec-Mspgf}

After a Euclidean rotation the magnetic spherical Green's function
$g^H_l(r,r^\prime;i\zeta)$ of Eq.\,(\ref{EM-GfunH-sp})
satisfies the differential equation
\begin{equation}
\left[ -\frac{1}{r} \frac{\partial}{\partial r} r 
\frac{1}{\varepsilon^\perp(r)} 
\frac{1}{r} \frac{\partial}{\partial r} r 
+ \frac{l(l+1)}{r^2\varepsilon^{||}(r)}
+ \zeta^2 \mu^\perp(r) \right] g^H_l(r,r^\prime;i\zeta)
= \frac{\delta(r-r^\prime)}{r^2}. 
\label{sradTMde}
\end{equation}
The continuity conditions on the spherical radial magnetic Green's functions 
are obtained by translating the boundary conditions on the fields 
given by Eqs.~(\ref{bc1-sp}) and (\ref{bc2-sp})
into the components of the Green's dyadics
in Eqs.~(\ref{E=GP-def}) and (\ref{H=GP-def}),
which are given in terms of the 
magnetic and electric spherical Green's functions using Eq.\,(\ref{G=XgX-sp}).
For two dielectric mediums separated by a spherical $\delta$-function sphere,
described by
\begin{subequations}
\begin{align}
\varepsilon^{\perp,{||}}(r) &= 1 + 
(\varepsilon^{\perp,{||}}_< -1) \,\theta(a-r)
+ (\varepsilon^{\perp,{||}}_> -1) \,\theta(r-a)
+ \lambda^{\perp,{||}}_e \,\delta (r-a), \\
\mu^{\perp,{||}}(r) &= 1 + (\mu^{\perp,{||}}_< -1) \,\theta(a-r)
+ (\mu^{\perp,{||}}_> -1) \,\theta(r-a) 
+ \lambda^{\perp,{||}}_g \,\delta (r-a),
\end{align}%
\label{epmu-def-sp}%
\end{subequations}
illustrated in Fig.~\ref{Dielectric-ball-fig}, the continuity conditions 
for the spherical radial magnetic Green's functions are derived to be
\begin{subequations}
\begin{align}
\frac{1}{\varepsilon^\perp(r)} \frac{1}{r} \frac{\partial}{\partial r} r 
g^H_l(r,r^\prime;i\zeta) \bigg|_{r=a-}^{r=a+} 
&= \zeta^2 \lambda_g^\perp \,g^H_l(r,r^\prime;i\zeta) \bigg|_{r=a}, \\
g^H_l(r,r^\prime;i\zeta) \bigg|_{r=a-}^{r=a+}
&= \lambda_e^\perp \, \frac{1}{\varepsilon^\perp(r)} 
\frac{1}{r} \frac{\partial}{\partial r} r g^H_l(r,r^\prime;i\zeta) \bigg|_{r=a}.
\end{align}%
\label{sradTMbc}%
\end{subequations}
The solution to the spherical radial magnetic Green's function in
Eq.\,(\ref{sradTMde}) satisfying the boundary conditions in 
Eqs.~(\ref{sradTMbc}) is
\begin{align}
&g^H_l(r,r^\prime;i\zeta) = \nonumber \\
& \begin{cases}
\frac{2}{\pi} \bar\zeta_< \varepsilon^\perp_<
\left[\text{i}_{\bar l_<}(\bar\zeta_< r_<) \text{k}_{\bar l_<}(\bar\zeta_< r_>) 
+ \sigma^{\text{scatt},H}_{l,<>} \,
\text{i}_{\bar l_<}(\bar\zeta_< r) \text{i}_{\bar l_<}(\bar\zeta_< r^\prime) 
\right], \quad & r,r^\prime < a, \\[3mm]
\frac{2}{\pi} \bar\zeta_> \varepsilon^\perp_>
\left[\text{i}_{\bar l_>}(\bar\zeta_> r_<) \text{k}_{\bar l_>}(\bar\zeta_> r_>) 
+ \sigma^{\text{scatt},H}_{l,><} \,
\text{k}_{\bar l_>}(\bar\zeta_> r) \text{k}_{\bar l_>}(\bar\zeta_> r^\prime) 
\right], & a<r,r^\prime, \\[3mm]
\frac{2}{\pi} \bar\zeta_> \varepsilon^\perp_>
\,\sigma^{\text{abs},H}_{l,><} \,
\text{i}_{\bar l_<}(\bar\zeta_< r) \text{k}_{\bar l_>}(\bar\zeta_> r^\prime),
& r<a<r^\prime, \\[3mm]
\frac{2}{\pi} \bar\zeta_< \varepsilon^\perp_<
\,\sigma^{\text{abs},H}_{l,<>} \,
\text{k}_{\bar l_>}(\bar\zeta_> r) \text{i}_{\bar l_<}(\bar\zeta_< r^\prime),
& r^\prime<a<r,
\end{cases}
\end{align}
where the scattering and absorption coefficients are
\begin{subequations}
\begin{align}
\sigma^{\text{scatt},H}_{l,<>} &= - \frac{ \big[ 
{\bar\lambda}^e_> ({\bar\lambda}^e_< \bar{\text{k}}_< + \text{k}_<) 
  (\bar{\text{k}}_> - {\bar\lambda}^g_> \text{k}_>)
+ {\bar\lambda}^e_< (\bar{\text{k}}_< + {\bar\lambda}^g_< \text{k}_<) 
  ({\bar\lambda}^e_> \bar{\text{k}}_> - \text{k}_>) \big] } 
{ \big[ {\bar\lambda}^e_> ({\bar\lambda}^e_< \bar{\text{i}}_< + \text{i}_<) 
  (\bar{\text{k}}_> - {\bar\lambda}^g_> \text{k}_>)
+ {\bar\lambda}^e_< (\bar{\text{i}}_< + {\bar\lambda}^g_< \text{i}_<) 
  ({\bar\lambda}^e_> \bar{\text{k}}_> - \text{k}_>) \big] },
\\ \sigma^{\text{scatt},H}_{l,><} &= - \frac{ \left[ 
{\bar\lambda}^e_< (\bar{\text{i}}_< + {\bar\lambda}^g_< \text{i}_<) 
  ({\bar\lambda}^e_> \bar{\text{i}}_> - \text{i}_>)
+ {\bar\lambda}^e_> ({\bar\lambda}^e_< \bar{\text{i}}_< + \text{i}_<) 
  (\bar{\text{i}}_> - {\bar\lambda}^g_> \text{i}_>) \right] }
{ \left[ {\bar\lambda}^e_> ({\bar\lambda}^e_< \bar{\text{i}}_< + \text{i}_<) 
  (\bar{\text{k}}_> - {\bar\lambda}^g_> \text{k}_>)
+ {\bar\lambda}^e_< (\bar{\text{i}}_< + {\bar\lambda}^g_< \text{i}_<) 
  ({\bar\lambda}^e_> \bar{\text{k}}_> - \text{k}_>) \right] },
\\ \sigma^{\text{abs},H}_{l,><} &= \frac{ \left[ 
{\bar\lambda}^e_> (\bar{\text{i}}_> - {\bar\lambda}^g_> \text{i}_>) 
  ({\bar\lambda}^e_> \bar{\text{k}}_> - \text{k}_>)
- {\bar\lambda}^e_> ({\bar\lambda}^e_> \bar{\text{i}}_> - \text{i}_>) 
  (\bar{\text{k}}_> - {\bar\lambda}^g_> \text{k}_>) \right] }
{ \left[ {\bar\lambda}^e_> ({\bar\lambda}^e_< \bar{\text{i}}_< + \text{i}_<) 
  (\bar{\text{k}}_> - {\bar\lambda}^g_> \text{k}_>)
+ {\bar\lambda}^e_< (\bar{\text{i}}_< + {\bar\lambda}^g_< \text{i}_<) 
  ({\bar\lambda}^e_> \bar{\text{k}}_> - \text{k}_>) \right] },
\\ \sigma^{\text{abs},H}_{l,<>} &= \frac{ \left[ 
{\bar\lambda}^e_< ({\bar\lambda}^e_< \bar{\text{i}}_< + \text{i}_<) 
  (\bar{\text{k}}_< + {\bar\lambda}^g_< \text{k}_<)
- {\bar\lambda}^e_< (\bar{\text{i}}_< + {\bar\lambda}^g_< \text{i}_<) 
  ({\bar\lambda}^e_< \bar{\text{k}}_< + \text{k}_<) \right] }
{ \left[ {\bar\lambda}^e_> ({\bar\lambda}^e_< \bar{\text{i}}_< + \text{i}_<) 
  (\bar{\text{k}}_> - {\bar\lambda}^g_> \text{k}_>)
+ {\bar\lambda}^e_< (\bar{\text{i}}_< + {\bar\lambda}^g_< \text{i}_<) 
  ({\bar\lambda}^e_> \bar{\text{k}}_> - \text{k}_>) \right] }.
\end{align}%
\label{scatt-coefs}%
\end{subequations}
We used the notations
\begin{equation}
\bar l_\lessgtr (\bar l_\lessgtr +1) = l(l+1) \,
\frac{\varepsilon^\perp_\lessgtr}{\varepsilon^{||}_\lessgtr}, \quad \quad
\bar\zeta^2_\lessgtr = \zeta^2 \mu^\perp_\lessgtr \varepsilon^\perp_\lessgtr,
\end{equation}
and
\begin{equation}
{\bar\lambda}^e_\lessgtr = \frac{\zeta\lambda^\perp_e}{2}
\sqrt{\frac{\mu^\perp_\lessgtr}{\varepsilon^\perp_\lessgtr}},
\quad \quad {\bar\lambda}^g_\lessgtr = \frac{\zeta\lambda^\perp_g}{2}
\sqrt{\frac{\varepsilon^\perp_\lessgtr}{\mu^\perp_\lessgtr}},
\end{equation}
and the shorthand notations 
\begin{subequations}
\begin{align}
\text{i}_\lessgtr &= \text{i}_{\bar l_\lessgtr} (\bar\zeta_\lessgtr a), \\
\text{k}_\lessgtr &= \text{k}_{\bar l_\lessgtr} (\bar\zeta_\lessgtr a),
\end{align}
\end{subequations}
where $\text{i}_l(t)$ and $\text{k}_l(t)$ are modified spherical Bessel 
functions that are related to the modified Bessel functions by the relations
\begin{subequations}
\begin{align}
\text{i}_l(t) &= \sqrt{\frac{\pi}{2t}} I_{l+\frac{1}{2}}(t), \\
\text{k}_l(t) &= \sqrt{\frac{\pi}{2t}} K_{l+\frac{1}{2}}(t).
\end{align}
\end{subequations}
In particular $\text{i}_l(t) = \text{i}^{(1)}_l(t)$, 
the modified spherical Bessel function of the first kind, 
together with $\text{k}_l(t)$ are suitable
pair of solutions in the right half of the complex plane
\cite{NIST:DLMF,NIST:2010fm}.
We used bars to define the following operations on the 
modified spherical Bessel functions
\begin{subequations}
\begin{align}
\bar{\text{i}}_l(t) 
&= \bigg( \frac{1}{t} + \frac{\partial}{\partial t} \bigg) \text{i}_l(t), \\
\bar{\text{k}}_l(t) 
&= \bigg( \frac{1}{t} + \frac{\partial}{\partial t} \bigg) \text{k}_l(t).
\end{align}%
\label{bar-mBf}%
\end{subequations}%
In Eqs.~(\ref{scatt-coefs}), using the scattering interpretation,
the first symbol after comma in the subscript of $\sigma$ 
denotes the medium of origin of the (incident) beam and
the second symbol denotes the medium of the transmitted beam.
Thus, for example, $\sigma^{\text{scatt},H}_{l,><}$ is the 
TM scattering amplitude for the case when the incident beam strikes
the sphere from outside of the sphere and
$\sigma^{\text{abs},H}_{l,><}$ is the corresponding TM absorption coefficient.
The analogs for a planar interface are the reflection and transmission
coefficients, respectively. Using the Wronskian for the 
modified spherical Bessel functions,
\begin{equation}
\text{k}_l \text{i}_l^\prime -\text{i}_l \text{k}_l^\prime = \frac{\pi}{2t^2},
\end{equation}
where primes denote differentiation, we have used the related Wronskian 
\begin{equation}
\text{k}_l \bar{\text{i}}_l -\text{i}_l \bar{\text{k}}_l = \frac{\pi}{2t^2}.
\end{equation}
The electric Green's function $g^E_l(r,r^\prime;i\zeta)$
is obtained from the magnetic Green's function $g^H_l(r,r^\prime;i\zeta)$
by swapping $\varepsilon \leftrightarrow \mu$ and $e \leftrightarrow g$ 
everywhere.

\subsubsection{Free Green's dyadic in spherical polar coordinates}

\begin{center} \underline{ 
($\varepsilon^{\perp,{||}}_\lessgtr=1$, $\mu^{\perp,{||}}_\lessgtr=1$,
and $\lambda^\perp_{e,g}=0$) }
\end{center}

After observing that
\begin{equation}
\left( \frac{1}{r} + \frac{\partial}{\partial r} \right)
\left( \frac{1}{r} + \frac{\partial}{\partial r} \right)
= \frac{1}{r^2} \frac{\partial}{\partial r} r^2 \frac{\partial}{\partial r},
\end{equation}
the free radial Green's function in vacuum is obtained by setting
$\varepsilon^{\perp,{||}}(r)=1$ and $\mu^{\perp,{||}}(r)=1$
in Eq.\,(\ref{sradTMde}) to yield
\begin{equation}
\left[ 
- \frac{1}{r^2} \frac{\partial}{\partial r} r^2 \frac{\partial}{\partial r} 
+ \frac{l(l+1)}{r^2} + \zeta^2 \right] g^{(0)}_l(r,r^\prime;i\zeta)
= \frac{\delta(r-r^\prime)}{r^2},
\label{sradfree}
\end{equation}
whose solution in terms of modified spherical Bessel functions is 
\begin{equation}
g^{(0)}_l(r,r^\prime;i\zeta) = \frac{2\zeta}{\pi} \,
\text{i}_l(\zeta r_<) \text{k}_l(\zeta r_>).
\label{fgf-sp}
\end{equation}
The free Green's dyadics in spherical coordinates are
\begin{align}
&\gamma_{lm}^{ij\,(0)} (r,r^\prime;\omega) = \nonumber \\
&\frac{2\zeta}{\pi} 
\left[ \begin{array}{c}
\zeta^2 \bar{\text{i}}_l(\zeta r_<) \bar{\text{k}}_l(\zeta r_>)
\hspace{20mm} 0 \hspace{10mm}
ik_\perp^\prime \zeta \left\{ \begin{array}{c} 
\bar{\text{i}}_l(\zeta r) \text{k}_l(\zeta r^\prime) \\[1mm]
\text{i}_l(\zeta r^\prime) \bar{\text{k}}_l(\zeta r) 
\end{array} \right\} \\[5mm] 
0 \hspace{15mm}
- \zeta^2 \text{i}_l(\zeta r_<) \text{k}_l(\zeta r_>)
\hspace{15mm} 0 \\[5mm]
-ik_\perp\zeta \left\{ \begin{array}{c} 
\text{i}_l(\zeta r) \bar{\text{k}}_l(\zeta r^\prime) \\[1mm]
\bar{\text{i}}_l(\zeta r^\prime) \text{k}_l(\zeta r) 
\end{array} \right\}
\hspace{8mm} 0 \hspace{12mm}
k_\perp k_\perp^\prime \text{i}_l(\zeta r_<) \text{k}_l(\zeta r_>)
\end{array} \right],
\end{align}
and
\begin{align}
& \phi_{lm}^{ij\,(0)} (r,r^\prime;\omega) = \nonumber \\
&-\frac{2\zeta^2}{\pi}
\left[ \begin{array}{ccc}
0 \hspace{13mm} 
\zeta \left\{ \begin{array}{c} 
\bar{\text{i}}_l(\zeta r) \text{k}_l(\zeta r^\prime) \\[1mm]
\text{i}_l(\zeta r^\prime) \bar{\text{k}}_l(\zeta r) 
\end{array} \right\} 
\hspace{13mm} 0 \\[5mm] 
\zeta \left\{ \begin{array}{c} 
\text{i}_l(\zeta r) \bar{\text{k}}_l(\zeta r^\prime) \\[1mm]
\bar{\text{i}}_l(\zeta r^\prime) \text{k}_l(\zeta r) 
\end{array} \right\} 
\hspace{12mm} 0 \hspace{12mm} 
ik_\perp^\prime \text{i}_l(\zeta r_<) \text{k}_l(\zeta r_>)
\\[5mm]
0 \hspace{12mm} 
-ik_\perp \text{i}_l(\zeta r_<) \text{k}_l(\zeta r_>)
\hspace{12mm} 0 \end{array} \right],
\end{align}
for $\left\{ \begin{array}{c} r < r^\prime \\ 
r^\prime < r \end{array} \right\}$.

\subsubsection{Magneto-electric ball in another magneto-electric medium}

\begin{center} \underline{
($\lambda^\perp_e=0$ and $\lambda^\perp_g=0$) }
\end{center}

Coefficients in Eqs.~(\ref{scatt-coefs}) for this case take the form
\begin{subequations}
\begin{align}
\sigma^{\text{scatt},H}_{l,<>} &= - \frac{ \left[
\sqrt{\frac{\mu^\perp_>}{\varepsilon^\perp_>}} \,\text{k}_< \bar{\text{k}}_>
- \sqrt{\frac{\mu^\perp_<}{\varepsilon^\perp_<}} \,\bar{\text{k}}_< \text{k}_>
\right] }{ \left[
\sqrt{\frac{\mu^\perp_>}{\varepsilon^\perp_>}} \,\text{i}_< \bar{\text{k}}_>
- \sqrt{\frac{\mu^\perp_<}{\varepsilon^\perp_<}} \,\bar{\text{i}}_< \text{k}_>
\right] } \xrightarrow{\varepsilon^\perp_>\to\infty}
- \frac{\bar{\text{k}}_<}{\bar{\text{i}}_<}
\xrightarrow{\varepsilon^\perp_<=1}
- \frac{\bar{\text{k}}}{\bar{\text{i}}}, \\
\sigma^{\text{scatt},H}_{l,><} &= - \frac{ \left[
\sqrt{\frac{\mu^\perp_>}{\varepsilon^\perp_>}} \,\text{i}_< \bar{\text{i}}_>
- \sqrt{\frac{\mu^\perp_<}{\varepsilon^\perp_<}} \,\bar{\text{i}}_< \text{i}_>
\right] }{ \left[
\sqrt{\frac{\mu^\perp_>}{\varepsilon^\perp_>}} \,\text{i}_< \bar{\text{k}}_>
- \sqrt{\frac{\mu^\perp_<}{\varepsilon^\perp_<}} \,\bar{\text{i}}_< \text{k}_>
\right] }
\xrightarrow{\varepsilon^\perp_<\to\infty}
- \frac{\bar{\text{i}}_>}{\bar{\text{k}}_>}
\xrightarrow{\varepsilon^\perp_>=1}
- \frac{\bar{\text{i}}}{\bar{\text{k}}}, \\
\sigma^{\text{abs},H}_{l,><} &=
- \frac{ \dfrac{\pi}{2} \dfrac{1}{(\bar\zeta_> a)^2} 
\sqrt{\frac{\mu^\perp_>}{\varepsilon^\perp_>}} }{ \left[
\sqrt{\frac{\mu^\perp_>}{\varepsilon^\perp_>}} \,\text{i}_< \bar{\text{k}}_>
- \sqrt{\frac{\mu^\perp_<}{\varepsilon^\perp_<}} \,\bar{\text{i}}_< \text{k}_>
\right] }
\xrightarrow{\varepsilon^\perp_<\to\infty}
-\frac{\pi}{2} \frac{1}{(\bar\zeta_> a)^2}
\frac{1}{ \text{i}_< \bar{\text{k}}_> } = 0, \\
\sigma^{\text{abs},H}_{l,<>} &=
- \frac{ \dfrac{\pi}{2} \dfrac{1}{(\bar\zeta_< a)^2} 
\sqrt{\frac{\mu^\perp_<}{\varepsilon^\perp_<}} }{ \left[
\sqrt{\frac{\mu^\perp_>}{\varepsilon^\perp_>}} \,\text{i}_< \bar{\text{k}}_>
- \sqrt{\frac{\mu^\perp_<}{\varepsilon^\perp_<}} \,\bar{\text{i}}_< \text{k}_>
\right] }
\xrightarrow{\varepsilon^\perp_>\to\infty}
\frac{\pi}{2} \frac{1}{(\bar\zeta_< a)^2}
\frac{1 }{ \bar{\text{i}}_< \text{k}_> },
\label{limch}
\end{align}
\end{subequations}
where $\text{i} = \text{i}_l(\zeta a)$ and $\text{k} = \text{k}_l(\zeta a)$.

\subsubsection{Semitransparent electric $\delta$-function sphere} 

\begin{center} \underline{
($\varepsilon^{\perp,{||}}_\lessgtr=1$, $\mu^{\perp,{||}}_\lessgtr=1$, 
and $\lambda^\perp_g=0$) }
\end{center}

Coefficients in Eqs.~(\ref{scatt-coefs}) for this case take the form
\begin{subequations}
\begin{align}
\sigma^{\text{scatt},H}_{l,<>} &=
\frac{\zeta\lambda^\perp_e \, \bar{\text{k}} \bar{\text{k}}}
{ \left[ \frac{\pi}{2\zeta^2a^2}
- \zeta\lambda^\perp_e \, \bar{\text{i}} \bar{\text{k}} \right] }
\xrightarrow{\lambda^\perp_e\to\infty}
- \frac{\bar{\text{k}}}{\bar{\text{i}}}, \\
\sigma^{\text{scatt},H}_{l,><} &= 
\frac{\zeta\lambda^\perp_e \, \bar{\text{i}} \bar{\text{i}}}
{ \left[ \frac{\pi}{2\zeta^2a^2}
- \zeta\lambda^\perp_e \, \bar{\text{i}} \bar{\text{k}} \right] }
\xrightarrow{\lambda^\perp_e\to\infty}
- \frac{\bar{\text{i}}}{\bar{\text{k}}}, \\
\sigma^{\text{abs},H}_{l,><} =
\sigma^{\text{abs},H}_{l,<>} &=
\frac{\frac{\pi}{2\zeta^2a^2}}
{ \left[ \frac{\pi}{2\zeta^2a^2} 
- \zeta\lambda^\perp_e \, \bar{\text{i}} \bar{\text{k}} \right] }
\xrightarrow{\lambda^\perp_e\to\infty} 0.
\end{align}%
\label{emds-le-lg0}%
\end{subequations}

\subsubsection{Semitransparent magnetic $\delta$-function sphere}

\begin{center} \underline{
($\varepsilon^{\perp,{||}}_\lessgtr=1$, $\mu^{\perp,{||}}_\lessgtr=1$,
and $\lambda^\perp_e=0$) }
\end{center}

Coefficients in Eqs.~(\ref{scatt-coefs}) for this case take the form
\begin{subequations}
\begin{align}
\sigma^{\text{scatt},H}_{l,<>} &=
- \frac{\zeta\lambda^\perp_g \, \text{k} \text{k}}
{ \left[ \frac{\pi}{2\zeta^2a^2}
+ \zeta\lambda^\perp_g \, \text{i} \text{k} \right] }
\xrightarrow{\lambda^\perp_g\to\infty}
- \frac{\text{k}}{\text{i}}, \\
\sigma^{\text{scatt},H}_{l,><} &=
- \frac{\zeta\lambda^\perp_g \, \text{i} \text{i}}
{ \left[ \frac{\pi}{2\zeta^2a^2}
+ \zeta\lambda^\perp_g \, \text{i} \text{k} \right] }
\xrightarrow{\lambda^\perp_g\to\infty}
- \frac{\text{i}}{\text{k}}, \\
\sigma^{\text{abs},H}_{l,><} =
\sigma^{\text{abs},H}_{l,<>} &=
\frac{\frac{\pi}{2\zeta^2a^2}}
{ \left[ \frac{\pi}{2\zeta^2a^2} 
+ \zeta\lambda^\perp_g \, \text{i} \text{k} \right] }
\xrightarrow{\lambda^\perp_g\to\infty} 0.
\end{align}%
\label{emds-le0-lg}%
\end{subequations}

\subsubsection{Semitransparent magneto-electric $\delta$-function sphere}

\begin{center} \underline{
($\varepsilon^{\perp,{||}}_\lessgtr=1$ and $\mu^{\perp,{||}}_\lessgtr=1$) }
\end{center}

Coefficients in Eqs.~(\ref{scatt-coefs}) for this case take the form
\begin{subequations}
\begin{align}
\sigma^{\text{scatt},H}_{l,<>} &=
\frac{ \left[ \zeta \lambda^\perp_e \, \bar{\text{k}} \bar{\text{k}}
- \zeta \lambda^\perp_g \, \text{k} \text{k} \right] }
{ \left[ (1+ \frac{1}{4} \zeta^2 \lambda^\perp_e \lambda^\perp_g)
\frac{\pi}{2\zeta^2a^2}
- \zeta\lambda^\perp_e \, \bar{\text{i}} \bar{\text{k}} 
+ \zeta\lambda^\perp_g \, \text{i} \text{k} \right] }
\xrightarrow[\lambda^\perp_g\to\infty]{\lambda^\perp_e\to\infty} 0, \\
\sigma^{\text{scatt},H}_{l,><} &= 
\frac{ \left[ \zeta \lambda^\perp_e \, \bar{\text{i}} \bar{\text{i}}
- \zeta \lambda^\perp_g \, \text{i} \text{i} \right] }
{ \left[ (1+\frac{1}{4} \zeta^2 \lambda^\perp_e \lambda^\perp_g)
\frac{\pi}{2\zeta^2a^2}
- \zeta\lambda^\perp_e \, \bar{\text{i}} \bar{\text{k}} 
+ \zeta\lambda^\perp_g \, \text{i} \text{k} \right] }
\xrightarrow[\lambda^\perp_g\to\infty]{\lambda^\perp_e\to\infty} 0, \\
\sigma^{\text{abs},H}_{l,><} &=
\frac{ (1-\frac{1}{4}\zeta^2 \lambda^\perp_e \lambda^\perp_g)
\frac{\pi}{2\zeta^2a^2} }
{ \left[ (1+\frac{1}{4}\zeta^2 \lambda^\perp_e \lambda^\perp_g)
\frac{\pi}{2\zeta^2a^2}
- \zeta\lambda^\perp_e \, \bar{\text{i}} \bar{\text{k}} 
+ \zeta\lambda^\perp_g \, \text{i} \text{k} \right] }
\xrightarrow[\lambda^\perp_g\to\infty]{\lambda^\perp_e\to\infty} -1, \\
\sigma^{\text{abs},H}_{l,<>} &= \sigma^{\text{abs},H}_{l,><}.
\end{align}%
\label{emdslelg}%
\end{subequations}

\subsection{Casimir-Polder interaction energy}

Our framework has been kept very general up till now.
It is possible, in our set-up, 
to consider the system illustrated in Fig.~\ref{Dielectric-ball-fig}
of a $\delta$-function sphere sandwiched between two mediums.
But, to limit the extent of our discussion here, we limit ourselves to 
an electric $\delta$-function sphere in vacuum, described by
\begin{equation}
{\bm\varepsilon}({\bf r};\omega) - {\bf 1}
= {\bf 1}_\perp \lambda(\omega) \delta(r-a), 
\quad {\bf 1}_\perp = {\bf 1} -\hat{\bf r}\hat{\bf r}
\label{dfun-sp-edef}
\end{equation}
and ${\bm\lambda}_g(\omega)=0$.
In particular, we shall proceed to evaluate the interaction energy 
between a polarizable molecule
described by Eqs.\,(\ref{pol-atom-e1}) and (\ref{pmol-eis})
and an electric $\delta$-function sphere of Eq.\,(\ref{dfun-sp-edef}).
The three dyadics in Eq.\,(\ref{Ecpa}), 
${\bm\Gamma}$, ${\bm\chi}$, and ${\bm\Gamma}_0$,
presented as a single dyadic in Eq.\,(\ref{G=GCG0def}),
now has spherical symmetry and thus is isotropic. 
But, the polarizable molecule is still completely anisotropic.
A spherically symmetric dyadic can be expressed in terms of the 
spherical vector eigenfunctions as per the decomposition of
the Green dyadic in Eq.\,(\ref{Gamma=gE-sp}). Using this feature 
the Casimir-Polder energy between an anisotropically polarizable 
molecule positioned a distance $h$ from the center of a 
spherically symmetric electric $\delta$-function sphere of radius $a$
is given by the expression
\begin{equation}
E = -a^2 \int_{-\infty}^\infty d\zeta \sum_{l=0/1}^\infty \,
\gamma_l^{ii^\prime}(h,a;\omega) 
\lambda^{i^\prime j^\prime}(\omega)
\gamma_l^{j^\prime j\,(0)}(a,h;\omega) \alpha_l^{ji}(\omega),
\label{Ecp-dm12}
\end{equation}
where the Green's dyadic matrices are given in terms of the free Green's
function in Eq.\,(\ref{fgf-sp}) as
\begin{equation}
\gamma_l^{j^\prime j\,(0)}(a,h;\omega)
= \frac{2\zeta^3}{\pi} \left[ \begin{array}{c}
\bar{\text{i}}_l(\zeta a) \bar{\text{k}}_l(\zeta h)
\hspace{10mm} 0 \hspace{6mm}
\frac{i k_\perp^\prime}{\zeta}
\bar{\text{i}}_l(\zeta a) \text{k}_l(\zeta h) \\[2mm]
0 \hspace{10mm} -\text{i}_l(\zeta a) \text{k}_l(\zeta h) \hspace{10mm} 0 \\[2mm]
-\frac{i k_\perp}{\zeta} \text{i}_l(\zeta a) \bar{\text{k}}_l(\zeta h)
\hspace{5mm} 0 \hspace{5mm}
\frac{k_\perp k_\perp^\prime}{\zeta^2} \text{i}_l(\zeta a) \text{k}_l(\zeta h)
\end{array} \right],
\label{rfgf-sp}
\end{equation}
where we used the notations, overriding the notation in Eq.\,(\ref{kperp-def}),
\begin{equation}
k_\perp^2 = \frac{l(l+1)}{a^2}, \qquad
k_\perp^{\prime 2} = \frac{l(l+1)}{h^2}.
\end{equation}
The implementation of the initial value for the summation on index $l$
in Eq.\,(\ref{Ecp-dm12}) will be explicitly clarified after 
Eq.\,(\ref{af-axis-sp}).
The Green functions for a $\delta$-function sphere, $g_l^{H,E}(h,a)$, 
requires the evaluation of the Green function on the surface of 
the $\delta$-function sphere, which is deduced
as the average of the limiting values of the Green functions
while approaching the surface from the inside and outside,
and is determined to be
\begin{equation}
g_l^{H,E}(h,a) = \frac{\zeta}{\pi} \begin{cases}
\Big[ 1+ \sigma_{l,\, h<a}^{\text{abs},H,E} \Big]
\text{i}_l(\zeta h) \text{k}_l(\zeta a)
+\sigma_{l,\,h<a}^{\text{scatt},H,E} \text{i}_l(\zeta a) \text{i}_l(\zeta h),
& h<a, \\[4mm]
\Big[ 1+ \sigma_{l,\,h>a}^{\text{abs},H,E} \Big]
\text{i}_l(\zeta a) \text{k}_l(\zeta h)
+\sigma_{l,\, h>a}^{\text{scatt},H,E} \text{k}_l(\zeta a) \text{k}_l(\zeta h),
& h>a, \end{cases}%
\label{ghe-dsp}%
\end{equation}%
where we have rewritten the Green's functions here after releasing most
of the compact notation used earlier in Sec.~\ref{sec-Mspgf}
to incorporate the generality in the solutions there.
The scattering and absorption coefficients that appeared in Sec.~\ref{sec-Mspgf}
will be presented in this notation subsequently for $h>a$ and $h<a$. 
We further introduce the notation
\begin{align}
g_l^{H,E}(\bar r,r^\prime) &= 
\frac{1}{\zeta} \left( \frac{1}{r} + \frac{\partial}{\partial r} \right) 
g_l^{H,E}(r,r^\prime), 
\label{der-gHE}
\end{align}
and the corresponding notation for the primed coordinate,
which are obtained from the solutions in Eq.\,(\ref{ghe-dsp}) by 
replacing the modified Bessel functions containing the variable with the bar
by the respective barred modified Bessel functions of Eqs.\,(\ref{bar-mBf}).
In terms of the Green functions and the corresponding derivative operations
in Eq.\,(\ref{der-gHE}) the Green dyadic matrix for the Green dyadic 
in the presence of $\delta$-function sphere is given by
\begin{equation}
\gamma_{l}^{ii^\prime}(h,a;\omega)
= \zeta^2 \left[ \begin{array}{ccc}
g_l^H(\bar h,\bar a) & 0 & 
\frac{i k_\perp^\prime}{\zeta} g_l^H(\bar h, a) \\[2mm]
0 & - g_l^E(h,a) & 0 \\[2mm]
-\frac{i k_\perp}{\zeta} g_l^H(h, \bar a)
& 0 &  \frac{k_\perp k_\perp^\prime}{\zeta^2} g_l^H(h,a)
\end{array} \right].
\label{redgf-edp-sp}
\end{equation}
An electric $\delta$-function sphere, like a $\delta$-function plate,
does not allow polarizability in the direction transverse to the surface
of the sphere, as a consequence of the constraint in Eq.\,(\ref{bc-sp}).
Thus, the matrix $\lambda^{i^\prime j^\prime}(\omega)$ is still given
by Eq.\,(\ref{mc-ld12}) with the matrix summation in indices on $u,v,w,$
now corresponding to the spherical vector eigenfunctions.
The fourth matrix in Eq.\,(\ref{Ecp-dm12}) represents the polarizable
molecule in which the vector spherical eigenfunctions are conveniently absorbed.
Without any loss of generality we can assume the polarizable molecule
to be on the $z$ axis, corresponding to choosing $\theta=0$ and $\phi=0$
inside the vector spherical eigenfunctions. Thus, we have
\begin{equation}
\alpha_l^{ji}(\omega) = \sum_{m=-l}^l {\bf X}_{lm}^{(j)*}(0,0)
\cdot {\bm\alpha}(\omega) \cdot {\bf X}_{lm}^{(i)}(0,0),
\end{equation}
where the matrix structure is similar to the cylindrical counterpart
in Eq.\,(\ref{aji-gend3}) with the vector eigenfunctions now being 
${\bf U}_{lm}(0,0)$, ${\bf V}_{lm}(0,0)$, and ${\bf W}_{lm}(0,0)$.
In terms of the unit vector
\begin{equation}
\hat{\bf t} = \frac{1}{\sqrt{2}} (\hat{\bm\theta} + i\hat{\bm\phi})
\end{equation}
evaluated on the axis we have
\begin{subequations}
\begin{align}
{\bf U}_{lm}(0,0) &= i\sqrt{\dfrac{2l+1}{8\pi}} 
\Big[ \delta_{m,1} \,\hat{\bf t} -\delta_{m,-1} \,\hat{\bf t}^* \Big], \\
{\bf V}_{lm}(0,0) &= \sqrt{\dfrac{2l+1}{8\pi}} 
\Big[ \delta_{m,1} \,\hat{\bf t} +\delta_{m,-1} \,\hat{\bf t}^* \Big],
\end{align}%
\label{uvm=00-sp12}
\end{subequations}%
and
\begin{equation}
{\bf W}_{lm}(0,0) = \hat{\bf z} \, \delta_{m0} \sqrt{\dfrac{2l+1}{4\pi}}, 
\end{equation}
using 
\begin{equation}
Y_{lm}(0,0)= \delta_{m0} \sqrt{\dfrac{2l+1}{4\pi}}.
\end{equation}
Using Eqs.\,(\ref{uvm=00-sp12}) we can evaluate,
\begin{align}
\alpha_l^{ji}(\omega) &= \sum_{m=-l}^l
{\bf X}_{lm}^{(j)*}(0,0) \cdot {\bm\alpha}(\omega) \cdot {\bf X}_{lm}^{(i)}(0,0)
\nonumber \\ &=
\dfrac{2l+1}{4\pi} \left[ \begin{array}{ccc}
\frac{1}{2} \text{tr}\,{\bm\alpha}_\perp(\omega) & 0 & 0 \\ 0 & 
\frac{1}{2} \text{tr}\,{\bm\alpha}_\perp(\omega) & 0 \\ 0 & 0 &
\hat{\bf z} \cdot {\bm\alpha}(\omega) \cdot \hat{\bf z}
\end{array} \right],
\label{af-axis-sp}
\end{align}
where 
$\text{tr}\,{\bm\alpha}_\perp(\omega) 
= \text{tr}\,{\bm\alpha}(\omega) \cdot {\bf 1}_\perp$, with
${\bf 1}_\perp = {\bf 1}- \hat{\bf r} \hat{\bf r}$ 
is evaluated on the axis to yield 
${\bf 1}_\perp = {\bf 1}- \hat{\bf z} \hat{\bf z}$. 
Thus, in Eqs.\,(\ref{redgf-edp-sp}), (\ref{mc-ld12}),
(\ref{rfgf-sp}), and (\ref{af-axis-sp}),
we have the explicit forms for the matrices 
$\gamma_l^{ii^\prime}$, $\lambda^{i^\prime j^\prime}$,
$\gamma_l^{j^\prime j\,(0)}$, and $\alpha^{ji}$, respectively, required 
for the evaluation of the Casimir-Polder energy in Eq.\,(\ref{Ecp-dm12}).
The summation over index $l$ in Eq.\,(\ref{Ecp-dm12}) start from
$l=1$ for terms containing $\text{tr}\,{\bm\alpha}_\perp(\omega)$
because its evaluation involves ${\bf U}_{lm}$ and ${\bf V}_{lm}$,
while it starts from $l=0$ for the terms involving
$\hat{\bf z} \cdot {\bm\alpha}(\omega) \cdot \hat{\bf z}$.
We shall now proceed to evaluate the energy for the cases when the 
polarizable molecule is either outside $(h>a)$ or inside $(h<a)$ the
electric $\delta$-function sphere.

\subsection{Polarizable molecule outside a $\delta$-function sphere: ($h>a$)}

\begin{figure}%
\begin{center}
\includegraphics{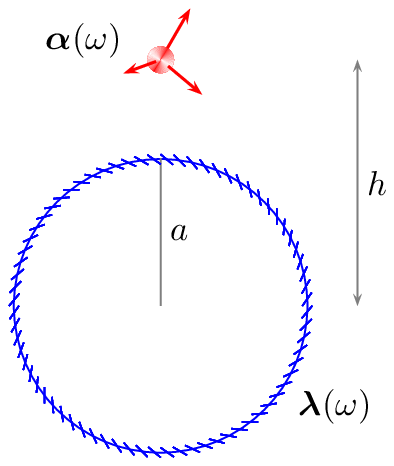}
\caption{An isotropically polarizable molecule of polarizability
${\bm\alpha}(\omega)$ outside a spherically symmetric
$\delta$-function sphere of radius $a$ described by ${\bm\lambda}(\omega)$.
The polarizable molecule is a distance $h>a$ from the center of 
the sphere.}%
\label{fig-mol-out-edsp}.%
\end{center}%
\end{figure}

The scattering coefficients for the case when the polarizable molecule
is outside the $\delta$-function sphere, $h>a$, 
as illustrated in Fig.~\ref{fig-mol-out-edsp}, are
\begin{subequations}
\begin{align}
\sigma_s^H \equiv \sigma^{\text{scatt},H}_{l,\,h>a} &=
\frac{\zeta\lambda \, \bar{\text{i}}_l(\zeta a) \bar{\text{i}}_l(\zeta a)}
{ \left[ \frac{\pi}{2\zeta^2a^2}
- \zeta\lambda \, \bar{\text{i}}_l(\zeta a) \bar{\text{k}}_l(\zeta a) \right] }
\xrightarrow{\lambda\to\infty}
- \frac{\bar{\text{i}}_l(\zeta a)}{\bar{\text{k}}_l(\zeta a)}, \\
\sigma_s^E \equiv \sigma^{\text{scatt},E}_{l,\,h>a} &=
- \frac{\zeta\lambda \, \text{i}_l(\zeta a) \text{i}_l(\zeta a)}
{ \left[ \frac{\pi}{2\zeta^2a^2}
+ \zeta\lambda \, \text{i}_l(\zeta a) \text{k}_l(\zeta a) \right] }
\xrightarrow{\lambda\to\infty}
- \frac{\text{i}_l(\zeta a)}{\text{k}_l(\zeta a)},
\end{align}%
\end{subequations}
and the absorption coefficients, which are independent of
whether the molecule is inside or outside, are
\begin{subequations}
\begin{align}
\sigma^{\text{abs},H}_l &=
\frac{\frac{\pi}{2\zeta^2a^2}}{ \left[ \frac{\pi}{2\zeta^2a^2}
- \zeta\lambda \, \bar{\text{i}}_l(\zeta a) \bar{\text{k}}_l(\zeta a) \right] }
\xrightarrow{\lambda\to\infty} 0, \\
\sigma^{\text{abs},E}_l &=
\frac{\frac{\pi}{2\zeta^2a^2}}{ \left[ \frac{\pi}{2\zeta^2a^2}
+ \zeta\lambda \, \text{i}_l(\zeta a) \text{k}_l(\zeta a) \right] }
\xrightarrow{\lambda\to\infty} 0.
\end{align}%
\label{ac-sp-lg0}%
\end{subequations}
The evaluation of the Casimir-Polder energy in Eq.\,(\ref{Ecp-dm12}) 
involves the products
\begin{subequations}
\begin{align}
\lambda \;g_l^H(\bar h,\bar a) 
\,\bar{\text{i}}_l(\zeta a) \bar{\text{k}}_l(\zeta h)
&= \frac{\sigma_s^H}{\zeta^2 a^2} \, \bar{\text{k}}_l^2(\zeta h), \\
\lambda \;g_l^E(h,a) 
\,\text{i}_l(\zeta a) \text{k}_l(\zeta h)
&= -\frac{\sigma_s^E}{\zeta^2 a^2} \, \text{k}_l^2(\zeta h), \\
\lambda \;g_l^H(h,\bar a) 
\,\bar{\text{i}}_l(\zeta a) \text{k}_l(\zeta h)
&= \frac{\sigma_s^H}{\zeta^2 a^2} \, \text{k}_l^2(\zeta h),
\end{align}
\end{subequations}
in terms of which the Casimir-Polder interaction energy of 
Eq.\,(\ref{Ecp-dm12}) between an anisotropically
polarizable molecule outside an electric $\delta$-function sphere
is given by the expression
\begin{align}
E &= - \int_{-\infty}^\infty \frac{d\zeta}{\pi} \zeta^3
\sum_{l=0/1}^\infty \frac{(2l+1)}{4\pi} 
\bigg[ \Big( \sigma_s^H \bar{\text{k}}_l^2(\zeta h)
-\sigma_s^E \text{k}_l^2(\zeta h) \Big) \,\text{tr}\,{\bm\alpha}(\omega)
\nonumber \\ & \hspace{5mm}
+\left( 2 \frac{l(l+1)}{\zeta^2 ah}
\sigma_s^H \text{k}_l^2(\zeta h)
-\sigma_s^H \bar{\text{k}}_l^2(\zeta h) +\sigma_s^E \text{k}_l^2(\zeta h) 
\right) 
\,\hat{\bf z} \cdot {\bm\alpha}(\omega) \cdot \hat{\bf z} \bigg],
\label{CP-sp-genE}
\end{align}
where we have used
\begin{equation}
\text{tr}\,{\bm\alpha}(\omega)
= \text{tr}\,{\bm\alpha}_\perp(\omega)
+ \hat{\bf z} \cdot {\bm\alpha}(\omega) \cdot \hat{\bf z}
\end{equation}
to bring out the correspondence between energy in Eq.\,(\ref{CP-sp-genE}) 
for a molecule interacting with a sphere and the energy in 
Eq.\,(\ref{CP-pl-genEan}) for a molecule interacting with a plane. 

In the perfect conductor limit the energy in Eq.(\ref{CP-sp-genE})
takes the form, expressed here in terms of 
$\text{tr}\,{\bm\alpha}_\perp(\omega)$,
\begin{equation}
E = - \int_{-\infty}^\infty \frac{d\zeta}{\pi} \zeta^3
\sum_{l=0/1}^\infty \frac{(2l+1)}{4\pi}
\bigg[ A_l(\zeta a,\zeta h) \text{tr}\,{\bm\alpha}_\perp(\omega)
+B_l(\zeta a,\zeta h) 
\,\hat{\bf z} \cdot {\bm\alpha}(\omega) \cdot \hat{\bf z} \bigg],
\label{CP-sp-conE}
\end{equation}
where
\begin{subequations}
\begin{align}
A_l(x,y) &=
-\frac{\bar{\text{i}}_l(x)}{\bar{\text{k}}_l(x)} \bar{\text{k}}_l^2(y)
+\frac{\text{i}_l(x)}{\text{k}_l(x)} \text{k}_l^2(y), \\
B_l(x,y) &= -2 l(l+1) \frac{1}{xy}
\frac{\bar{\text{i}}_l(x)}{\bar{\text{k}}_l(x)} \text{k}_l^2(y).
\end{align}
\end{subequations}
Further, in the Casimir-Polder limit, where only the static polarizabilities
are relevant, in addition to the perfect conductor limit,
the energy can be expressed in the form
\begin{equation}
E = -\frac{\text{tr}\,{\bm\alpha}_\perp(0)}{8\pi (h-a)^4} 
\,f\left( \frac{h}{a} \right)
-\frac{\hat{\bf z} \cdot {\bm\alpha}(0) \cdot \hat{\bf z}}{8\pi (h-a)^4} 
\,g\left( \frac{h}{a} \right),
\end{equation}
where
\begin{subequations}
\begin{align}
f\left( \frac{h}{a} \right) &= \frac{4}{\pi} \int_0^\infty y^3 dy
\sum_{l=1}^\infty (2l+1) A_l\left(\frac{a\,y}{h-a},\frac{h\,y}{h-a}\right)
\xrightarrow{h\to a} 1, \\
g\left( \frac{h}{a} \right) &= \frac{4}{\pi} \int_0^\infty y^3 dy
\sum_{l=0}^\infty (2l+1) B_l\left(\frac{a\,y}{h-a},\frac{h\,y}{h-a}\right)
\xrightarrow{h\to a} 0.
\end{align}%
\label{fgha-def}%
\end{subequations}%
We remind that the index $l$ begins from $l=1$ for $f(h/a)$,
because it is built out of vector spherical eigenfunctions
${\bf U}_{lm}$ and ${\bf V}_{lm}$, while it starts from $l=0$
for $g(h/a)$ because it is built out of ${\bf W}_{lm}$.
The functions $f(h/a)$ and $g(h/a)$ are chosen such that the
interaction energy of a sphere reduces to that of a plate in the 
limit $h\to a$. Unlike in the case of the plate, for a sphere
the interaction is sensitive to the orientation of the polarizabilities
of the molecule even in the perfect conductor limit.
We shall investigate the orientation dependence in 
the energy in Eq.\,(\ref{CP-sp-genE}) elsewhere. 

Within the confines of the Casimir-Polder limit and 
the perfect conductor limit
let us further narrow down our attention to the special case when the
polarizable molecule is unidirectionally polarizable. 
In addition, to consider an axially symmetric configuration,
let us place the molecule on the $z$ axis with it's 
direction of polarizability aligned with the $z$ axis, such that we have
\begin{equation} 
{\bm\alpha}(\omega) = \alpha(\omega)\, \hat{\bf z} \hat{\bf z}. 
\label{alzzsp}
\end{equation}
For this, axially symmetric, scenario the interaction energy 
in Eq.(\ref{CP-sp-genE}) takes the form
\begin{equation}
E = - \frac{\alpha(0)}{8\pi (h-a)^4} g\left( \frac{h}{a} \right).
\label{cpe-sp-asym}
\end{equation}

\subsection{Polarizable molecule inside a $\delta$-function sphere: ($h<a$)}

\begin{figure}%
\begin{center}
\includegraphics{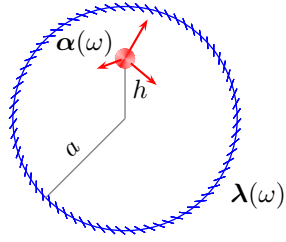}
\caption{An isotropically polarizable molecule of polarizability
${\bm\alpha}(\omega)$ inside a spherically symmetric
$\delta$-function sphere of radius $a$ described by ${\bm\lambda}(\omega)$.
The polarizable molecule is a distance $h<a$ from the center of      
the sphere.}%
\label{fig-mol-in-edsp}.%
\end{center}%
\end{figure}

The scattering coefficients for the case when the polarizable molecule
is inside the $\delta$-function sphere, $h>a$,
illustrated Fig.~\ref{fig-mol-in-edsp}, are
\begin{subequations}
\begin{align}
\sigma^{\text{scatt},H}_{l,\,h<a} &=
\frac{\zeta\lambda \, \bar{\text{k}}_l(\zeta a) \bar{\text{k}}_l(\zeta a)}
{ \left[ \frac{\pi}{2\zeta^2a^2} - \zeta\lambda \, 
\bar{\text{i}}_l(\zeta a) \bar{\text{k}}_l(\zeta a) \right] }
\xrightarrow{\lambda\to\infty}
- \frac{\bar{\text{k}}_l(\zeta a)}{\bar{\text{i}}_l(\zeta a)}, \\
\sigma^{\text{scatt},E}_{l,\,h<a} &=
\frac{\zeta\lambda \, \text{k}_l(\zeta a) \text{k}_l(\zeta a)}
{ \left[ \frac{\pi}{2\zeta^2a^2} - \zeta\lambda \,
\text{i}_l(\zeta a) \text{k}_l(\zeta a) \right] }
\xrightarrow{\lambda\to\infty}
- \frac{\text{k}_l(\zeta a)}{\text{i}_l(\zeta a)},
\end{align}%
\end{subequations}%
and the absorption coefficients, independent of whether the 
molecule is inside or outside the sphere, is given by 
Eqs.\,(\ref{ac-sp-lg0}).

The Casimir-Polder interaction energy of
Eq.\,(\ref{Ecp-dm12}) between an anisotropically
polarizable molecule and an electric $\delta$-function sphere,
when the molecule is inside the sphere, is obtained from the
energy for the molecule outside the sphere by swapping
the modified spherical Bessel functions 
$\text{i}_l \leftrightarrow \text{k}_l$
everywhere, including the barred Bessel functions,
inside the expression for energy in Eq.\,(\ref{CP-sp-genE}).
For the special case when the polarizable molecule is at the center
of the perfectly conducting sphere, obtained for $h=0$,
the interaction energy in the Casimir-Polder limit has the form
\begin{equation}
E = -\frac{\text{tr}\,{\bm\alpha}_\perp(0)}{8\pi a^4} \,f(0),
\end{equation}
where $f(0)$ is transcribed from Eqs.\,(\ref{fgha-def})
by swapping the modified spherical Bessel functions,
\begin{equation}
f(0) = \frac{4}{\pi} \int_0^\infty x^3 dx
\sum_{l=1}^\infty (2l+1) \left[
\frac{\text{k}_l(x)}{\text{i}_l(x)} \text{i}_l^2(0)
-\frac{\bar{\text{k}}_l(x)}{\bar{\text{i}}_l(x)} \bar{\text{i}}_l^2(0) \right].
\end{equation}
Using the asymptotic form for zero argument of modified spherical
Bessel functions we have $\text{i}_l(0)=1$, if $l=0$, and 0, otherwise.
Further, $\bar{\text{i}}_l(0)=2/3$, if $l=1$, and 0, for $l>1$.
Thus, we have
\begin{equation}
f(0) = -\frac{16}{3\pi} \int_0^\infty x^3 dx
\, \frac{\bar{\text{k}}_1(x)}{\bar{\text{i}}_1(x)} \sim 8.48049.
\end{equation}
The term in the energy containing 
$\hat{\bf z} \cdot {\bm\alpha}(0) \cdot \hat{\bf z}$
does not appear in the expression of energy when $h=0$ because
\begin{equation}
g(0) = -\frac{8}{\pi} \int_0^\infty x^3 dx
\sum_{l=0}^\infty (2l+1) \frac{l(l+1)}{x^2}
\frac{\bar{\text{k}}_l(x)}{\bar{\text{i}}_l(x)}
\lim_{z\to 0} \frac{\text{i}_l^2(z)}{z} =0,
\end{equation}
using $\lim_{z\to 0} \text{i}_l^2(z)/z=0$, if $l\neq 0$, and since $l=0$ term
is zero explicitly. 
When the molecule is at the center of a perfectly conducting sphere
the energy is expected to be insensitive to the orientation of the 
polarizabilities of the molecule.
The apparent dependence on only two of the principal polarizabilities,
of the three, is misleading, because we chose the polarizable molecule to be 
positioned on the $z$ axis. 

\section{Casimir-Polder interaction energy: Proposition}
\label{sec-thesis}

To follow up on the proposition in Sec.~\ref{sec-cp-prop},
let us consider an unidirectionally polarizable molecule described by
Eqs.\,(\ref{pol-atom-e1}) and (\ref{alzzsp}) interacting with a 
$\delta$-function spheroid described by
\begin{equation}
{\bm\chi}({\bf r};\omega)
= {\bf 1}_\perp \frac{\lambda(\omega)}{c} \delta(\xi-\xi_2), \qquad 
{\bf 1}_\perp = \hat{\bm\eta} \hat{\bm\eta} + \hat{\bm\phi} \hat{\bm\phi},
\end{equation}
in terms of spheroidal coordinates~\cite{NIST:DLMF,NIST:2010fm} 
$(\xi,\eta,\phi)$ and positive constant $c$, 
the interaction energy for which is given by
\begin{equation}
E = -\alpha(0) \lambda(0) \int_{-\infty}^\infty d\zeta \int d^2 r_\perp
\hat{\bf z} \cdot {\bm\Gamma}({\bf r}_0,{\bf r}_\perp)
\cdot {\bf 1}_\perp \cdot
{\bm\Gamma}_0({\bf r}_\perp,{\bf r}_0) \cdot \hat{\bf z}.
\label{cpe-spd}
\end{equation}
The Green dyadics of Eq.\,(\ref{cpe-spd}) have decompositions
of the form similar to the ones in  Eqs.\,(\ref{G=XgX-sp}), but now
in terms of spheroidal vector eigenfunctions ${\bf X}^{(i)}_{lm}(\eta,\phi)$.
Unlike for the case of a plate and a sphere, now the reduced Green's
functions are dependent on the $m$ mode and the interaction energy is given by
\begin{equation}
E=-\int_{-\infty}^\infty d\zeta \sum_{m=0}^\infty \sum_{l=m}^\infty
\gamma^{ii^\prime}_{lm}(\xi_1,\xi_2;\omega) \lambda^{i^\prime j^\prime}(\omega)
\gamma^{j^\prime j\;(0)}_{lm}(\xi_2,\xi_1;\omega) 
\alpha^{ji}_{lm}(\omega),
\end{equation}
where $\xi_1$ and $\xi_2$ specify the position of the molecule 
and the surface of the spheroid, respectively.
The matrix elements of the polarizability of the molecule is 
given in terms of the angular spheroidal vector eigenfunctions
${\bf W}_{lm}(\eta,\phi)$ as
\begin{equation}
\alpha^{ji}_{lm}(\omega) = {\bf W}_{lm}^*(1,0) 
\cdot \hat{\bf z}\, \alpha(\omega)\, \hat{\bf z} \cdot {\bf W}_{lm}(1,0)
= d \,\delta_{m0} \alpha(\omega),
\end{equation}
where $d$ is a constant given in terms of the product of angular
(oblate) spheroidal functions, 
the spheroidal functions of the first kind~\cite{NIST:DLMF,NIST:2010fm},
$\text{Ps}_l^m(\eta,\zeta^2a^2)$, evaluated at $\eta=1$.
This renders the energy to get non-zero contribution only from the $m=0$ mode.
And, that is not all, there is a bonus.
In general, the transverse electric (TE) and transverse magnetic (TM) modes
do not separate in spheroidal coordinates, but, the $m=0$ azimuth mode
does indeed separate into TE and TM modes. Our proposition is that this is
a generic feature for axially symmetric systems, that is, only the $m=0$ 
azimuth mode contributes to the Casimir-Polder interaction energy.
The result for the interaction energy between an
unidirectionally polarizable molecule and a $\delta$-function spheroid, 
when the molecule is placed on the axis of symmetry of a oblate spheroid 
such that the direction of polarizability is aligned with the axis, 
is given by an expression similar to Eq.\,(\ref{cpe-sp-asym}) obtained by the 
replacing modified spherical Bessel functions with radial
spheroidal functions~\cite{NIST:DLMF,NIST:2010fm}, $m=0$,
\begin{subequations}
\begin{align}
\text{i}_l(\zeta r) &\to S_l^{m\,(1)}(i\xi,\zeta c), \\
\text{k}_l(\zeta r) &\to S_l^{m\,(2)}(i\xi,\zeta c).
\end{align}
\end{subequations}
We shall present the details of the calculation for the spheroid
in a sequel.

\section{Conclusions}

We have developed a formalism suitable for studying electrodynamics
in the presence of a medium having the symmetry of a coordinate surface.
It involves the decomposition of vector fields in the basis of 
vector eigenfunctions specific to the coordinate surface. 
The decomposition facilitates a systematic inspection of the separation
of electric and magnetic fields into TE and TM modes. 
Even though we have not presented the details for any
non-trivial coordinate surfaces here, we have put forward the
proposition that the $m=0$ azimuth mode for axially symmetric surfaces
always separates into TE and TM modes. 
We have further argued that the Casimir-Polder interaction
energy between an axially symmetric dielectric body and an
unidirectionally polarizable molecule, when the molecule is positioned
on the symmetry axis of the dielectric body with its polarizability 
in the direction of the axis, 
gets non-zero contribution only from the $m=0$ azimuth mode. 
Together, the suggestion is that Casimir-Polder energies for 
axially symmetric configurations can be easily evaluated.

As an application of the methods developed here, we have derived the
Casimir-Polder energy between an anisotropically polarizable molecule
and an electric $\delta$-function surface, the surfaces considered
being a plate and a sphere. We have also outlined a procedure that can
be used to derive the energy when the coordinate surface is a spheroid.
We believe, these methods can be extended for a half hyperboloid,
which for spheroidal coordinate $\eta=0$ corresponds to a plate
with an aperture. We intend to develop and present these results
in a follow-up paper.

\section*{Acknowledgements}

We thank Kim Milton for useful comments during the course of
this ongoing program, and Simen A. Ellingsen for collaborative 
assistance during the early stages of the project. 
We acknowledge support from the Research Council of Norway 
(Project No. 250346).

\appendix
\renewcommand*{\thesection}{\Alph{section}}
\section{Axial symmetry}
\label{sec-ax-sym-pg}

A rotation is unambiguously described by a vector ${\bm\omega}$, that is,
the angle representing the amount of rotation is contained in the 
magnitude of the vector, $\omega$, and the axis of rotation
is represented by the direction $\hat{\bm\omega}$ of the vector.
The transformation of a position vector ${\bf r}$ under
an infinitesimal rotation $\delta{\bm\omega}$ is given by 
\begin{equation}
\bar{\bf r} = {\bf r} - \delta{\bf r}, \qquad
\delta{\bf r} = \delta{\bm\omega} \times{\bf r},
\end{equation}
in the passive description.
In particular, a rotation about the direction $\hat{\bf z}$
by an infinitesimal (azimuth) angle $\delta\phi$ is described by
\begin{equation}
\delta{\bm\omega} = \hat{\bf z}\,\delta\phi.
\label{zphi-def}
\end{equation} 
The corresponding infinitesimal transformation in ${\bf r}$ is given by
\begin{equation}
\delta{\bf r} = \delta\omega\, \hat{\bf z} \times{\bf r}
= \hat{\bm\phi}\,\rho\delta\phi,
\label{rvar=zp}
\end{equation}
where $\rho$ and $\phi$ are the cylindrical coordinates defined as
\begin{equation}
\hat{\bf z}\times \hat{\bf r} = \hat{\bm\phi}
\quad \text{and} \quad
|\hat{\bf z}\times {\bf r}| =\rho.
\end{equation}
Observe that, in rectangular coordinates
$\rho\hat{\bm\phi} = x\hat{\bf y} -y\hat{\bf x}$.

A (scalar, vector, or tensor,) field ${\bf M}({\bf r})$ has
axial (azimuthal) symmetry about the $\hat{\bf z}$ axis
if it does not change under azimuthal variations of $\delta\phi$ 
in Eqs.\,(\ref{zphi-def}) and (\ref{rvar=zp}). That is,
the response of the field, given by
\begin{equation}
{\bf M}({\bf r}) \to {\bf M}({\bf r}) +\delta {\bf M}({\bf r}),
\end{equation}
satisfies the variational statement
\begin{equation}
\frac{\delta {\bf M}}{\delta\phi} =0.
\label{gen-axial-def}
\end{equation}

The transformation of a scalar field $\Phi({\bf r})$ 
under infinitesimal rotation is given by~\cite{Schwinger:1989psf1}
\begin{equation}
\delta\Phi({\bf r}) = \delta{\bm\omega} \cdot ({\bf r} \times {\bm\nabla}) \Phi,
\label{sc-rot-def}
\end{equation}
where ${\bm\nabla}$ is the gradient operator.
A scalar field with axial symmetry about the $\hat{\bf z}$ axis satisfies,
using Eq.\,(\ref{zphi-def}) in Eq.\,(\ref{sc-rot-def}),
\begin{equation}
\frac{\delta\Phi}{\delta\phi} 
= \hat{\bf z}\cdot({\bf r} \times{\bm\nabla})\Phi
= \frac{\partial\Phi}{\partial\phi} =0,
\label{sc-axial-def}
\end{equation}
where we used Eq.\,(\ref{rvar=zp}) and the cylindrical coordinate
representation of gradient to write the second equality.
The constraint imposed by Eq.\,(\ref{sc-axial-def}) is the statement
of axial symmetry for a scalar field as per Eq\,(\ref{gen-axial-def}).

The transformation of a vector field ${\bf E}({\bf r})$               
under infinitesimal rotation is given by~\cite{Schwinger:1989psf1}
\begin{equation}
\delta{\bf E}({\bf r}) 
=\delta{\bm\omega} \cdot ({\bf r} \times{\bm\nabla}) {\bf E}
-\delta {\bm\omega} \times {\bf E}.
\label{vec-rot-def}
\end{equation}
A vector field with axial symmetry about the $\hat{\bf z}$ axis satisfies,
using Eq.\,(\ref{zphi-def}) in Eq.\,(\ref{vec-rot-def}),
\begin{equation}
\frac{\delta{\bf E}}{\delta\phi} 
=\hat{\bf z} \cdot ({\bf r} \times{\bm\nabla}) {\bf E} 
-\hat{\bf z} \times {\bf E} = 0.
\label{vec-axial-def}
\end{equation}
Using 
\begin{equation}
\hat{\bf z} \cdot ({\bf r} \times{\bm\nabla}) = \frac{\partial}{\partial\phi}
\end{equation}
and the completeness relation for the basis vectors in cylindrical 
coordinates,
\begin{equation}
\hat{\bf z} \hat{\bf z} + \hat{\bm\rho} \hat{\bm\rho}
+ \hat{\bm\phi} \hat{\bm\phi} = {\bf 1},
\end{equation}
to write
\begin{equation}
\hat{\bf z} \times {\bf E} = \hat{\bf z} \times {\bf 1} \cdot {\bf E}
=(\hat{\bm\phi} \hat{\bm\rho} - \hat{\bm\rho} \hat{\bm\phi}) \cdot {\bf E},
\end{equation}
we can rewrite the statement of axial symmetry for vector field in
Eq.\,(\ref{vec-axial-def}) as
\begin{equation}
\frac{\delta{\bf E}}{\delta\phi} = \frac{\partial {\bf E}}{\partial\phi}
+ (\hat{\bm\rho} \hat{\bm\phi} -\hat{\bm\phi} \hat{\bm\rho}) \cdot {\bf E} =0.
\label{vec-as-01}
\end{equation}
In the representation
\begin{equation}
{\bf E}({\bf r}) = \hat{\bm\rho} \,E_\rho({\bf r})
+ \hat{\bm\phi} \,E_\phi({\bf r}) + \hat{\bf z} \,E_z({\bf r})
\label{vec-as-def12}
\end{equation}
the statement of Eq.\,(\ref{vec-as-01}) concerning
axial symmetry of a vector field reads
\begin{equation}
\frac{\delta{\bf E}}{\delta\phi} 
= \hat{\bm\rho} \frac{\partial E_\rho}{\partial\phi}
+ \hat{\bm\phi} \frac{\partial E_\phi}{\partial\phi}
+ \hat{\bf z} \frac{\partial E_z}{\partial\phi} =0,
\end{equation}
which implies that the projection of the vector field along all
the basis vectors is independent of the azimuth angle $\phi$,
\begin{equation}
\frac{\partial E_\rho}{\partial\phi} =0, \quad
\frac{\partial E_\phi}{\partial\phi} =0, \quad
\frac{\partial E_z}{\partial\phi} =0.
\label{vec-com-as}
\end{equation}
The independence in $\phi$ of the components of an axially symmetric
vector field given in Eq.\,(\ref{vec-com-as}) can be also expressed as
\begin{equation}
\frac{\partial}{\partial\phi} 
\Big[ \hat{\bf e}_i \cdot {\bf E}({\bf r}) \Big] =0,
\quad \hat{\bf e}_1 =\hat{\bf z}, \quad \hat{\bf e}_2 =\hat{\bm\rho},
\quad \hat{\bf e}_3 =\hat{\bm\phi}.
\label{vec-com-asc}
\end{equation}

As an example of an axially symmetric field we have
${\bf E}({\bf r}) = \hat{\bm\rho}$,
corresponding to $E_\rho=1$, $E_\phi=0$, and $E_z=0$.
Another example is ${\bf E}({\bf r}) = \hat{\bm\phi}$,
corresponding to $E_\rho=0$, $E_\phi=1$, and $E_z=0$. Thus,
using Eq.\,(\ref{vec-as-def12}),
\begin{subequations}
\begin{align}
\frac{\delta\hat{\bm\rho}}{\delta\phi}
&= \frac{\partial\hat{\bm\rho}}{\partial\phi} 
-\hat{\bf z} \times \hat{\bm\rho} =0, \\
\frac{\delta\hat{\bm\phi}}{\delta\phi}
&= \frac{\partial\hat{\bm\phi}}{\partial\phi} 
-\hat{\bf z} \times \hat{\bm\phi}=0.
\end{align}%
\end{subequations}%
Further, ${\bf E}({\bf r}) = \hat{\bf z}$,
corresponding to $E_\rho=0$, $E_\phi=0$, and $E_z=1$,
is also an axially symmetric field.
The distinction between vectors and vector fields in these examples
is exemplified by contrasting the above variations with the vector identities
\begin{equation}
\frac{\partial \hat{\bm\rho}}{\partial\phi} =\hat{\bm\phi}, \quad
\frac{\partial \hat{\bm\phi}}{\partial\phi} =-\hat{\bm\rho}, \quad
\frac{\partial \hat{\bf z}}{\partial\phi} =0.
\end{equation}

Similarly, the transformation of a dyadic field ${\bf M}({\bf r})$
under infinitesimal rotation is given by~\cite{Schwinger:1989psf1}
\begin{equation}
\delta{\bf M}({\bf r})
=\delta{\bm\omega} \cdot ({\bf r} \times{\bm\nabla}) {\bf M}
-\delta {\bm\omega} \times {\bf M} +{\bf M} \times\delta {\bm\omega},
\label{dy-rot-def}
\end{equation}
and an axially symmetric dyadic field corresponds to 
all its components being independent of the azimuth angle $\phi$,
\begin{equation}
\frac{\partial}{\partial\phi}
\Big[ \hat{\bf e}_i \cdot {\bf M}({\bf r}) \cdot \hat{\bf e}_j \Big] =0,
\end{equation}
for $\hat{\bf e}_{i,j}$'s defined in Eq.\,(\ref{vec-com-asc}).

\subsection{Green's dyadic}

A scalar Green's function $G({\bf r},{\bf r}^\prime)$ depends on two
positions, ${\bf r}$ and ${\bf r}^\prime$, often referred to as position
of source and observer. A scalar Green's function, in our study here,
will be related to the vacuum expectation value of the product of two
scalar fields by the correspondence
\begin{equation}
\frac{1}{i} G({\bf r},{\bf r}^\prime)
=\langle \Phi({\bf r}) \Phi({\bf r}^\prime) \rangle.
\end{equation}
Under an infinitesimal rotation $\delta{\bm\omega}$ the transformation
of a scalar Green's function is dictated by the product structure of fields
Using Eq.\,(\ref{sc-rot-def}) for the individual fields in the product
structure we deduce that
\begin{equation}
\delta G({\bf r},{\bf r}^\prime)
= \delta{\bm\omega} \cdot ({\bf r} \times {\bm\nabla}) G 
+ \delta{\bm\omega} \cdot ({\bf r}^\prime \times {\bm\nabla}^\prime) G. 
\label{sc-gf-rot}
\end{equation}
An axially symmetric scalar Green's function thus satisfies,
using Eqs.\,(\ref{sc-axial-def}) and (\ref{sc-gf-rot}),
\begin{equation}
\frac{\delta G}{\delta\phi} =\left( 
\frac{\partial}{\partial\phi} +\frac{\partial}{\partial\phi^\prime} \right)
G({\bf r},{\bf r}^\prime) =0.
\end{equation}

If the dependence in the two position ${\bf r}$ and ${\bf r}^\prime$ in
the scalar Green's is of the form $|{\bf r}-{\bf r}^\prime|$, then, using
\begin{equation}
|{\bf r}-{\bf r}^\prime|
=\sqrt{
r^2 +{r^\prime}^2 -2rr^\prime
\big\{\cos\theta \cos\theta^\prime
+\sin\theta \sin\theta^\prime \cos(\phi-\phi^\prime) \big\} }
\end{equation}
and
\begin{equation} 
\left( \frac{\partial}{\partial\phi} 
+\frac{\partial}{\partial\phi^\prime} \right) \cos(\phi-\phi^\prime) =0,
\end{equation}
we deduce the axial symmetry of such scalar Green's functions.
Another way a scalar Green's function could be axially symmetric
is when the source point ${\bf r}^\prime$ is on the symmetry axis 
$\hat{\bf z}$ of rotation, that is,
\begin{equation}
{\bf r}^\prime = \hat{\bf z}\,h,
\end{equation}
because then we have the response
\begin{equation}
\delta{\bm\omega} \times {\bf r}^\prime
= \delta\phi\, \hat{\bf z} \times \hat{\bf z}\,h =0.
\end{equation}
Thus, in this scenario, for axial symmetry, it is sufficient if the 
dependence of the scalar Green's function is independent of $\phi$, that is,
\begin{equation}
\frac{\delta G}{\delta\phi} 
= \frac{\partial}{\partial\phi} G({\bf r},{\bf r}^\prime) =0.
\end{equation}

Our study in this article will be centered around the Green's dyadic
${\bm\Gamma}({\bf r},{\bf r}^\prime)$ which is related to the
vacuum expectation value of the product of two vector fields by the 
correspondence
\begin{equation}
\frac{1}{i} {\bm\Gamma}({\bf r},{\bf r}^\prime)
=\langle {\bf E}({\bf r}) {\bf E}({\bf r}^\prime) \rangle.
\end{equation}
The transformation of a Green's dyadic under an infinitesimal rotation 
$\delta{\bm\omega}$ is given by
\begin{equation}
\delta {\bm\Gamma}({\bf r},{\bf r}^\prime)
= \delta{\bm\omega} \cdot ({\bf r} \times {\bm\nabla}) {\bm\Gamma}
+ \delta{\bm\omega} \cdot ({\bf r}^\prime \times {\bm\nabla}^\prime){\bm\Gamma} 
-\delta{\bm\omega} \times {\bm\Gamma}
+{\bm\Gamma} \times \delta {\bm\omega}.
\label{gdya-rot}
\end{equation}

The Green's dyadic for the vacuum, which is termed the free Green's dyadic,
satisfies the dyadic differential equation
\begin{equation}
-\left[ \frac{1}{\zeta^2} {\bm\nabla} \times {\bm\nabla} \times +{\bf 1} \right]
\cdot {\bm\Gamma}_0({\bf r},{\bf r}^\prime) 
= {\bf 1} \delta^{(3)} ({\bf r}-{\bf r}^\prime),
\end{equation}
which has a formal solution of the form
\begin{equation}
{\bm\Gamma}_0({\bf r},{\bf r}^\prime) 
= \big[ {\bm\nabla} {\bm\nabla} -\zeta^2 {\bf 1} \big]
G_0(|{\bf r}_0-{\bf r}^\prime|). 
\end{equation}
For an infinitesimal rotation about the $\hat{\bf z}$ axis we have
\begin{align}
\frac{\delta{\bm\Gamma}_0}{\delta\phi} &= \left( 
\frac{\partial}{\partial\phi} +\frac{\partial}{\partial\phi^\prime} \right)
{\bm\Gamma}_0({\bf r},{\bf r}^\prime) 
-\hat{\bf z} \times {\bm\Gamma}
+{\bm\Gamma} \times \hat{\bf z} \nonumber \\
&= \big[ {\bm\nabla} {\bm\nabla} -\zeta^2 {\bf 1} \big]
\left( \frac{\partial}{\partial\phi} +\frac{\partial}{\partial\phi^\prime} 
\right) G_0(|{\bf r}_0-{\bf r}^\prime|) =0,
\end{align}
using 
\begin{equation}
\frac{\partial}{\partial\phi} {\bm\nabla}G
=\hat{\bf z} \times {\bm\nabla} G + {\bm\nabla} \frac{\partial}{\partial\phi}G.
\end{equation}
Thus, the free Green's dyadic is axially symmetric.

\bibliographystyle{elsarticle-num}
\section*{References}
\bibliography{%
biblio/b20120508-EM-modes-axial,%
}

\begin{thebibliography}{10}
\expandafter\ifx\csname url\endcsname\relax
  \def\url#1{\texttt{#1}}\fi
\expandafter\ifx\csname urlprefix\endcsname\relax\def\urlprefix{URL }\fi
\expandafter\ifx\csname href\endcsname\relax
  \def\href#1#2{#2} \def\path#1{#1}\fi

\bibitem{Morse:1953mp}
P.~M. Morse, H.~Feshbach, Methods of Theoretical Physics, McGraw-Hill Book
  Company, New York, 1953.

\bibitem{Stratton:1941em}
J.~A. Stratton, Electromagnetic Theory, McGraw-Hill Book Company, New York,
  1941.

\bibitem{Schwinger:1998cla}
J.~Schwinger, L.~L. DeRaad, Jr., K.~A. Milton, W.-y. Tsai,
  \href{https://archive.org/details/ClassicalElectrodynamicsSchwingerDeraadMiltonTsai}{Classical
  electrodynamics}, Advanced book program, Perseus Books, 1998.

\bibitem{Parashar:2012it}
P.~Parashar, K.~A. Milton, K.~V. Shajesh, M.~Schaden,
  \href{https://doi.org/10.1103/PhysRevD.86.085021} {Electromagnetic
  semitransparent $\delta$-function plate: Casimir interaction energy between
  parallel infinitesimally thin plates}, Phys. Rev. D 86 (2012) 085021.

\bibitem{Milton:2013bm}
K.~A. Milton, P.~Parashar, M.~Schaden, K.~V. Shajesh,
  \href{https://doi.org/10.1393/ncc/i2013-11532-4} {Casimir interaction
  energies for magneto-electric $\delta$-function plates}, Nuovo Cim. C036~(03)
  (2013) 193.

\bibitem{Schwinger:1977pa}
J.~Schwinger, L.~L. DeRaad, Jr., K.~A. Milton,
  \href{https://doi.org/10.1016/0003-4916(78)90172-0} {Casimir effect in
  dielectrics}, Annals Phys. 115 (1978) 1.

\bibitem{Thiyam:2015fts}
P.~Thiyam, P.~Parashar, K.~V. Shajesh, C.~Persson, M.~Schaden, I.~Brevik, D.~F.
  Parsons, K.~A. Milton, O.~I. Malyi, M.~Bostr\"om,
  \href{http://dx.doi.org/10.1103/PhysRevA.92.052704} {Anisotropic contribution
  to the van der {W}aals and the {C}asimir-{P}older energies for
  ${\text{CO}}_{2}$ and ${\text{CH}}_{4}$ molecules near surfaces and thin
  films}, Phys. Rev. A 92 (2015) 052704.

\bibitem{Casmir:1947hx}
H.~B.~G. Casimir, D.~Polder, \href{http://dx.doi.org/10.1103/PhysRev.73.360}
  {The Influence of retardation on the London-van der Waals forces}, Phys. Rev.
  73 (1948) 360.

\bibitem{Parashar:2017sgo}
P.~Parashar, K.~A. Milton, K.~V. Shajesh, I.~Brevik,
  \href{https://arxiv.org/abs/1708.01222} {Electromagnetic $\delta$-function
  sphere}, arXiv:1708.01222 [hep-th].

\bibitem{NIST:DLMF}
\href{http://dlmf.nist.gov/} {NIST Digital Library of Mathematical Functions},
  Release 1.0.8 of 2014-04-25, online companion to \cite{NIST:2010fm}.

\bibitem{NIST:2010fm}
\href{http://dlmf.nist.gov/} {{NIST} handbook of mathematical functions},
  Cambridge University Press, New York, 2010, edited by F. W. J. Olver and D.
  W. Lozier and R. F. Boisvert and C. W. Clark. Print companion to
  \cite{NIST:DLMF}.

\bibitem{Schwinger:1989psf1}
J.~Schwinger, Particles, Sources, and Fields, Volume I, Addison-Wesley,
  Massachusetts, 1970.

\end{thebibliography}


\end{document}